\documentclass[fleqn,twoside]{article}
\usepackage[headings]{espcrc2}

\readRCS
$Id: espcrc2.tex,v 1.2 2004/02/24 11:22:11 spepping Exp $
\usepackage{graphicx}
\usepackage[figuresright]{rotating}

\newcommand{\AmS}{{\protect\the\textfont2
  A\kern-.1667em\lower.5ex\hbox{M}\kern-.125emS}}
  
\title{Optical Properties of\\III-Mn-V Ferromagnetic Semiconductors}
\author{K.S. Burch\address[UT]{Department of Physics and Institute for Optical Sciences, University of Toronto, Toronto, Ontario, M5S 1A7 Canada}$^{,}$\thanks{kburch@physics.utoronto.ca}, D.D. Awschalom\address[UCSB]{Department of Physics, University of California, Santa Barbara, CA 93106}, D.N. Basov\address[UCSD]{Department of Physics, University of California, San Diego, CA 92093-0319}}

\runtitle{Optical Properties of III-Mn-V Ferromagnetic Semiconductors}
\runauthor{K.S. Burch}

\begin{document}

\begin{abstract}
We review the first decade of extensive optical studies of ferromagnetic, III-Mn-V diluted magnetic semiconductors. Mn introduces holes and local moments to the III-V host, which can result in carrier mediated ferromagnetism in these disordered semiconductors. Spectroscopic experiments provide direct access to the strength and nature of the exchange between holes and local moments; the degree of itineracy of the carriers; and the evolution of the states at the Fermi energy with doping. Taken together, diversity of optical methods reveal that Mn is an unconventional dopant, in that the metal to insulator transition is governed by the strength of the hybridization between Mn and its p-nictogen neighbor. The interplay between the optical, electronic and magnetic properties of III-Mn-V magnetic semiconductors is of fundamental interest and may enable future spin-optoelectronic devices.
 \end{abstract}

\maketitle

\section{Introduction}
	For more than two decades the study of magnetic impurities in semiconductors has received a great deal of attention (see \cite{Schneider,furdyna_review,DietlReview,samarth_review,OhnoH:ProfIs,Clerjaud:1985lr,jungwirth:809} and references therein). The interest in these compounds is driven in part by the desire to utilize the spin degree of freedom in electronics\cite{WolfSA:Spisev,ZuticI:Spifaa,Awschalom:2007lr}.  This rapidly expanding field of spin based electronics (spintronics) has exploited metal based devices to bring about a revolution in magnetic sensors and computer hard drive density\cite{Grochowski}. The discovery of ferromagnetism in III-V hosts heavily doped with Mn has launched a new phase in magnetic semiconductors research in part because of high promise of III-Mn-V systems in the context of Spintronics applications. The unique advantage of III-Mn-VÕs is that that magnetic, optical and electronic effects are all interconnected. Furthermore, all these properties are sensitive to external stimuli and can be readily modified by the application of external electric or magnetic field and also by illumination with light. Inherent tunablity of (ferro)magnetic semiconductors is of high interest in view of novel magneto-optical device functionalities. Since optical techniques are well suited to create, measure and manipulate spins in semiconductors \cite{WeberCP:ObssCd,stephens:097602,KikkawaJM:SpicsS,SalisG:Elecsc,KatoYK:Obstsh,CrookerSA:Imastl,ZakJ:Uniam} the systematic investigation of optical properties of III-Mn-V materials promises to uncover new avenues for applications of ferromagnetic semiconductors by offering insights into the fundamental physical processes in these systems.  
			
	Magnetic semiconductors also pose a number of unique challenges to understanding the physics governing their properties. In addition to the ``standard'' difficulties with separating electron-electron interactions from disorder effects,\cite{SarachikReview} magnetic dopants result in additional states in the band structure whose description is non-trivial\cite{HaldaneTMSemiconductors}. Nonetheless they also provide a rare opportunity to study the connections between band structure, carrier dynamics and magnetism in a well-controlled environment. Initially this work focused on II-VI compounds and very dilute concentrations of transition metals (TM) doped into III-V semiconductors\cite{Schneider,furdyna_review,Brandt_review,DietlBook,Clerjaud:1985lr}. Research on these diluted magnetic semiconductors (DMS) was limited by the difficulty in doping carriers into II-VI compounds and the low concentration limit for TM in III-V materials\cite{Clerjaud:1985lr}. The origins of the limit in III-V semiconductors lies in the different lattice structures for the end members, namely zinc-blende or wurzite for III-V compounds and hexagonal for TM-V materials\cite{ZungerGaP}. Nonetheless this limitation has been overcome via growth under non-equilibrium conditions,\cite{Munekata_InMnAs_first,OhnoGMSFirst} resulting in a wide variety of III-TM-V compounds and heterostructures (including superlattices and quantum wells) that exhibit magnetic order and can be tuned through their metal to insulator transition (MIT). 
	
	This advance has opened up a large array of possible applications for ferromagnetic III-Mn-V materials and led to an enormous amount of research. Despite the substantial progress over the past decade, there are still a number of open questions in this field. In particular, the proper theoretical framework to describe the ferromagnetism that emerges in many III-TM-V materials still remains elusive\cite{jungwirth:809,TimmReview,mahadevan:115211,zhang:125303,picozzi:235207,InMnAs_cr_thry,alvarez:045202,SanvitoGaMnAsPrb,fernandez-rossier:127201,lee-2007,zhou:144419,majidi:115205,tang:047201,DietlReview,fiete:045212,hwang:035210}. The answer to this fundamental question has been hindered by a lack of consensus about the evolution of the electronic structure of III-TM-V compounds with TM doping. A goal of this review is to offer the reader an in-depth discussion of this fundamental, yet controversial issue. In this vain, we have based our discussion on the information that has recently emerged from a variety of spectroscopic experiments with an emphasis on data generated through extensive optical studies of  III-Mn-V materials.

\begin{figure}
\center
\includegraphics[width=18pc]{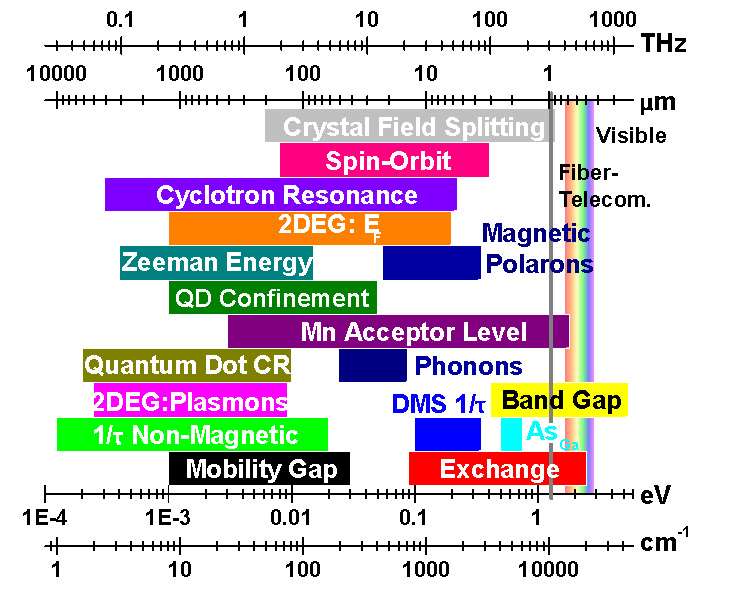}
\caption{\label{fig:energyScales} Energy scales in semiconductors covered by optical techniques, with $\frac{1}{\tau}$ the free carrier scattering rate, $E_{F}$ the Fermi energy. The abbreviations QD, CR, DMS, and 2DEG refer to quantum dot, cyclotron resonance, diluted magnetic semiconductor, and 2-Dimensional Electron Gas, respectively.}
\end{figure}

	Optical techniques have established themselves as providing key insights into the band structure of materials. In fact, the electronic structure of many semiconductors has been determined by comparing theoretical calculations to optical results\cite{CardonaBook}. Furthermore, as summarized in Figure \ref{fig:energyScales}, optical techniques cover a broad range of energy scales in magnetic semiconductors. In addition, optical studies have provided a unique view into the underlying physics governing various correlated electron materials\cite{BasovReview,Degiorgi,TokuraReview,ImadaMITReview,Dordevic_review}. Similar to these systems III-TM-V semiconductors also exhibit an interplay between their magnetic properties, carrier dynamics and band structure. Interestingly in III-TM-V DMS the carriers originate from the magnetic ions but also mediate the long range magnetic interaction by traversing the host lattice. This arises from the mixing of the local moments with extended states. In fact, this hybridization between p and d orbitals is what distinguishes III-TM-V compounds from semiconductors doped with non magnetic impurities. 
	
	To better understand the effects of magnetism on the band structure of III-V semiconductors (and vice versa) extensive optical studies have been applied to III-Mn-V ferromagnetic DMS. As detailed throughout this review, magneto-optical experiments have demonstrated the ferromagnetism originates from a single III-Mn-V phase and not from inclusions, such as Mn-V in a III-V matrix. Furthermore, magneto-optical measurements have provided a determination of the strength of the hybridization between the Mn local moments and holes that mediate the magnetic state\cite{AndoDerivMCD,DietlBook,DietlReview,furdyna_review,KojiAndo06302006}. As far as the evolution of the electronic structure with Mn doping is concerned, optical spectroscopy has uncovered a number of fundamental distinctions with the properties triggered by non-magnetic impurities in the III-V series.  In particular, the electromagnetic response of some heavily doped III-Mn-Vs that reveal a metallic state are in stark contrast to the optical properties of non magnetic counterparts. It is our intention to present a brief overview of the investigation of the optical properties of III-Mn-V semiconductors at the time of the completion of this paper. The review is intended to help the readers, especially those who are not spectroscopy experts, to navigate through the vast literature as well as to learn about both resolved and outstanding problems.

\subsection{Theoretical Background}
One of the earliest successes of quantum mechanics was the explanation of the role of the band structure in separating metals from insulators. There are many examples in the literature where a material is tuned from one of these diametrically opposed states to the other. Understanding metal to insulator transitions continues to be a central challenge not only in the study of doped semiconductors, but to many areas of contemporary condensed matter physics\cite{SarachikReview,MottBook,ImadaMITReview,efrosBook,LeeReviewMIT,ThomasReviewMIT}. Since the long-range magnetic order in magnetic semiconductors often emerges close to the metallic state, a complete description of the MIT appears important to fully understanding the ferromagnetism in III-Mn-V semiconductors. We therefore begin by discussing our current understanding of the MIT of semiconductors doped with non-magnetic, charged impurities. The discussion will focus on the implications for the electronic structure of traversing a metal to insulator transition, as well as what role the band structure plays in the MIT. This section is meant to simply give the reader a basic overview of this complex topic, a more complete discussion can be found in references \cite{ImadaMITReview,MottBook,efrosBook,LeeReviewMIT,ThomasReviewMIT}.

\begin{figure}
\center
\includegraphics[width=18pc]{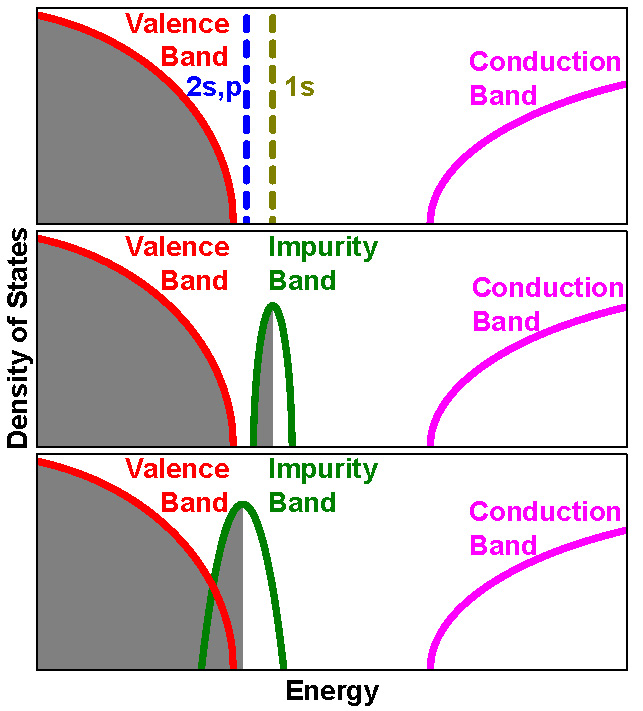}
\caption{\label{fig:IBGraph} A diagram of the evolution of the electronic structure in a hole doped semiconductor. Top panel: Two hydrogenic states are shown for a single acceptor with the 1s level set by the binding energy of an electron in a hydrogen atom, re-normilized by the dielectric constant and effective mass of the host. Middle panel: At higher doping levels the hole states broaden into an impurity band. Bottom panel: With additional doping the impurity band broadens from the increased overlap of the holes and acceptors. Furthermore the screening of the Coulomb attraction of the holes to the acceptors results in the impurity band moving closer to the valence band.}
\end{figure}

	Let us begin by considering the effect of doping a single charged impurity, namely an acceptor that introduces a hole into the system. Since the acceptor will gain one electron from bonding with its neighbors, the hole will experience a Coulomb attraction to the acceptor, thereby behaving like a hydrogen atom. However since the semiconductor is highly polarizable, it will screen the charge significantly. As a result, new states form in the band gap of the semiconductor that correspond to the states of a bound electron in a hydrogen atom whose energies have been reduced by a factor of $(\frac{m_{VB}}{m_{e}})\frac{1}{\epsilon^2}$, where $m_{VB}$ is the effective mass of holes in the valence band, $m_{e}$ is the free electron mass and $\epsilon$ is the static dielectric constant. In fact, a whole series of states will form, that correspond to the screened states of a hydrogen atom, often referred to as the Lyman series and labeled as in atomic spectra (1s, 2s, 2p etc...).

	At very low doping levels and in the limit $T\rightarrow0$, the holes donated by the acceptors occupy hydrogenic  bound states. This means the doped semiconductor will still be an insulator (ie: zero d.c. conductivity, $\sigma_{dc}$, in the limit of zero temperature). Early in the study of doped semiconductors, Mott pointed out that these states are modified as the dopant density is increased. With additional overlap the states will start to broaden into an ``impurity band" (IB)\cite{MottNF:thetmc}.  Initially the IB will be near the 1s state of the original acceptor and its width will be defined, in part, by the average overlap of the holes on different acceptor sites (See Fig \ref{fig:IBGraph}). Since, generally speaking, there is no exchange splitting in non-magnetic semiconductors the IB will be half-filled, therefore one might expect metallic behavior as soon as an IB is well formed. However the separation of the impurity band  from the VB indicates that there is still some Coulomb attraction between the holes and the acceptors. This implies that at zero temperature a bound state can always be formed and the material remains an insulator. Furthermore the Coulomb repulsion between the holes as well as the disorder inevitable in any doped system will likely lead to localization of the hole states. 
	
	The metallic transport in doped semiconductors is believed to occur via the assumption that once the Coulomb attraction between the holes and the acceptors is completely screened the IB ``dissolves" into the main band. This implies that the holes now occupy Bloch states and do not form bound states with the acceptors, resulting in metallic behavior. Within this scenario one might expect a ``minimum" metallic conductivity, as the carriers can now conduct charge as they would in a standard metal. Lastly we note that since the impurity band tends to be quite close to the main band, the conduction at high temperatures is enhanced via thermal activation. The experimental situation in real materials is often more complicated than this simplified description. For example, in conflict with the above scenario metallic transport has been observed to occur within an IB.\cite{DrewNTypeGaAs} However the existence of a metallic conduction in an impurity band has not been explained.

	We note that there are two important aspects of the MIT that we have mostly ignored. The first is the Coulomb repulsion between carriers, in opposition to the Coulomb attraction of the holes to the acceptors discussed above. Specifically, there is some energy cost associated with placing two holes on one acceptor site. This energy is generally referred to as the Hubbard ``U", and will tend to split the impurity band into upper and lower Hubbard bands, corresponding to singly and doubly occupied sites respectively\cite{MottBook,efrosBook,ThomasReviewMIT}. While this U term is usually quite large for electrons in d and f orbitals, in doped semiconductors the wave-functions of the holes tend to be quite extended. Therefore the splitting between the upper and lower Hubbard bands is typically quite small. Since this splitting is generally doping independent, one may expect a MIT when the bandwidth of the Hubbard bands becomes bigger than U. Such a transition is generally referred to as a Mott-Hubbard metal to insulator transition. However the large exchange splitting in III-Mn-V DMS excludes a Mott-Hubbard transition,\cite{ImadaMITReview,MottBook,ColemanLocalMoment} as explained in the next section. 
	
	The other important aspect of the metal to insulator transition in doped semiconductors, is the effect of disorder. Five decades ago Anderson showed that if the disorder was strong enough the carriers would become localized\cite{Andersonlocalization}. What Anderson realized is that disorder has the effect of placing an envelope on the wave-function of the carriers. This exponential decay will have a localization length that is proportional to the bandwidth of the band containing the carriers divided by strength of the disorder\cite{MottBook,efrosBook,LeeReviewMIT}. Therefore, although the carriers may not be in a bound state, if the localization length is smaller than the length of the sample, the carriers can no longer diffuse across the sample, preventing conduction at zero temperature. This in turn implies that as one tunes the strength of disorder, and/or the bandwidth, the localization length diverges, taking the sample from an insulating to a metallic state. Generally speaking it is the states in the tails of the band that are localized. The energy separating localized from itinerant states is called a mobility edge, and so as one tunes the Fermi energy, (for example by changing the degree of compensation), an Anderson MIT can occur as $E_{F}$ goes through the mobility edge. \cite{ImadaMITReview,MottBook,efrosBook,LeeReviewMIT}
	
\subsection{Magnetism and the Metal to Insulator Transition}
\label{sec:magMIT}		
	One might ask, how the MIT in doped semiconductors is affected by the presence of magnetic impurities. Noting that a full discourse on this topic is also quite complex, we will focus on the case of Mn in a III-V semiconductor. In general, Mn will replace the cation (group III element) in the lattice (we will refer to these substitutional atoms as $Mn_{III}$). When the Mn substitutes for the cation, it is nominally in the $Mn^{3+}$ state with 4 electrons in its outer d-shell, which we denote as $d^{4}$. Upon bonding with the anion, the Mn may absorb an electron converting it to $Mn^{2+}$ and producing a hole. Similar to a non-magnetic impurity the hole will experience a Coulomb attraction to the Mn creating an acceptor level. However in the case of Mn, the electron will end up in the outer d-shell, creating a $d^{5}$ local moment. Due to the strong U on these orbitals, this half-full shell will be spin-polarized. Since the hole is created by the transfer of charge from the group V element's p-orbital to the Mn d-orbital, the strength of the exchange between the hole and the local moment is proportional to the overlap of the these orbitals. This mixing of local Mn and extended group V element states is generally referred to as p-d hybridization ($V_{pd}$).\cite{jungwirth:809,DietlReview,HassReview}
	
	The exchange between the Mn local moments and the carriers they produce, plays a key role in the physics of III-Mn-V diluted magnetic semiconductors. In fact, it is this exchange that enables the carriers to mediate the ferromagnetism between the dilute moments. In DMS the exchange is characterized by two constants, $\alpha$ and $\beta$. These refer to the exchange coupling of the conduction ($\alpha$) and valence ($\beta$) bands to the local moments. The exchange occurs via two mechanisms Kinetic and Coulomb, where the latter results from virtual jumps between orbitals (ie: As to Mn and back to As). Therefore the sign and magnitude of $\alpha$ \& $\beta$ depends on the hybridization of the orbitals involved. Specifically, s-d hybridization is forbidden by symmetry at the $\Gamma$ point, therefore $\alpha$ is believed to originate from direct (ie: Coulomb) exchange between electrons and Mn.  

	The polarization of the holes that results from Kinetic exchange has a number of implications for the MIT in DMS. Specifically, since the holes are created in a spin-selective way the impurity band they form is not spin-degenerate.\cite{DietlReview} Therefore the Mott-Hubbard picture of the MIT is no longer relevant. Another important aspect of the p-d hybridization is its tendency to localize the holes around the Mn. This effect originates from the exchange energy gained by the hole by being in the vicinity of the Mn. Therefore in III-Mn-V DMS the effect of disorder tends to be stronger and the strength of the Anderson localization is influenced by the size of $V_{pd}$\cite{DietlReview,TimmReview}. 
	
	The tendency of the holes to localize around the Mn also has important implications for the electronic structure of the III-Mn-V DMS. Specifically the binding energy of the hole will not only include its Coulomb attraction to the Mn nucleus but also the strength of $V_{pd}$. Therefore the impurity band tends to be significantly further away from the valence band than in other p-type III-V semiconductors\cite{Bhattacharjee}.  Another interesting implication of the hybridization between the holes and the Mn local moment, is its effect on the nature of the states in the impurity band. In particular, it has been suggested that this impurity band will not originate only from the states taken from the valence band (primarily p-like states of the group V element), but the IB may also acquire d-orbital character from the Mn. The degree of  Mn character in the impurity band will be determined by the strength of $V_{pd}$\cite{alvarez:045202,SanvitoGaMnAsPrb,berciu:045202,MillisPRLGaMnAs,ErwinGaMnAsPRL,rader:075202}.  Since the impurity band is no longer simply derived from valence band states, this suggests that if the hybridization is strong enough the standard picture of an IB that merges back into the main band upon the MIT\cite{ImadaMITReview,MottBook,efrosBook,LeeReviewMIT} may need to be modified.  This concept can be checked by tuning the lattice constant (and in turn the hybridization between Mn and the group V element), while simultaneously monitoring the strength of the exchange and examining the electronic structure. As we will detail, optics has demonstrated that the exchange is stronger in materials with smaller lattice constants. Furthermore the impurity band appears to persist on the metallic side of the MIT in some of these compounds (see sub-section \ref{sec:GaMnAsIR}). Provided $V_{pd}$ is strong enough, then the Mn d states end up inside the band gap of the semiconductor, such that the hole remains in a Mn d-level. This implies that for strong enough $V_{pd}$, Mn remains in the d$^4$ configuration (see sections \ref{sec:GaMnNMagOpt} and \ref{sec:GaMnNLum}). 
	
\section{In$_{1-x}$Mn$_{x}$As}
\label{sec:InMnAs}	
	Munekata and collaborators were the first to overcome the apparent intrinsic doping limits of Mn in III-V semiconductors by growing In$_{1-x}$Mn$_{x}$As under non-equilibrium conditions using low temperature molecular beam epitaxy(LT-MBE)\cite{Munekata_InMnAs_first}. This quickly led to the first realization by Ohno \textit{et al.} of a ferromagnetic III-V DMS \cite{Ohno_FM_InMnAs}. It was also found that the samples could be prepared as n-type or p-type depending on growth conditions, however only the p-type films exhibited ferromagnetic behavior. InAs has the largest lattice constant of all of the III-Mn-V systems, and therefore should have the smallest value of exchange (see section \ref{sec:magMIT}). The ferromagnetic transition temperatures (T$_{C}$) of In$_{1-x}$Mn$_{x}$As compounds tend to be somewhat lower than in other III-Mn-Vs (i.e. T$_{C}< 90~K$).\cite{OhnoH:Molbea,Munekata90KInMnAs} 

\subsection{Cyclotron Resonance in In$_{1-x}$Mn$_{x}$As}	
\label{sec:CR}
	Since determining the nature of the holes in III-Mn-V DMS is a key to understanding the physics underlying their magnetic order, cyclotron resonance (CR) experiments emerged as an important probe of these compounds. In fact, the first CR experiments in semiconductors were also the first verification of the validity of the quasi-particle concept, which is a cornerstone of Fermi liquid theory. In these measurements, the microwave absorption was monitored in Ge\cite{Original_cyclotron}. In this case the carriers were optically excited into the conduction and valence bands and strong absorption was seen at the appropriate cyclotron frequencies ($\omega_{c}$). The cyclotron frequency depends directly on the band mass $m_{B}$ of the quasi-particles ($\omega_{c}=\frac{eB}{m_{B}c}$), where $e$ is the free electron charge, $B$ the applied magnetic field, and $c$ is the speed of light. Therefore by measuring the absorption at a fixed frequency and sweeping the magnetic field (note that the external radiation must propagate along the direction of applied magnetic field) one will observe an absorption at a B determined by $m_{B}$. Thus Dresslhaus et. al's observation of a cyclotron absorption at fields much higher than what would be expected for free electrons, demonstrated that the carriers in Ge and Si moved with a band mass significantly smaller than $m_{e}$. 
	
	It is interesting to note that despite the potential importance of cyclotron measurements to the understanding of  III-Mn-V magnetic semiconductors, these experiments were not reported until 2002. So far CR has only been observed in In$_{1-x}$Mn$_{x}$As. While this might seem surprising, one must keep in mind that for a CR peak to be observed the condition: $\omega_{c}>\frac{1}{\tau}$ must be satisfied, where $\frac{1}{\tau}$ is the scattering rate of the carriers. Since III-Mn-V materials contain significant numbers of defects, overcoming the $\omega_{c}>\frac{1}{\tau}$ condition requires very large magnetic fields (between 50 and 500 Tesla).  Another useful aspect of CR experiments is that the scattering rate can be determined from the width of the absorption peak. The CR of In$_{1-x}$Mn$_{x}$As was first measured in an n-type, paramagnetic sample\cite{InMnAs_ntype_cr}. In this experiment a resonant feature that shifts to lower B and broadens as the amount of Mn is increased was observed. These results indicated a reduction of the electron band mass (approximately $25\%$ for $x=0.12$) and increased disorder. While this study did not directly address the nature of the carriers that mediate the ferromagnetism in In$_{1-x}$Mn$_{x}$As, it was able to determine the value of the exchange through direct comparison with theoretical predictions. 

\begin{figure}
\center
\includegraphics[width=18pc]{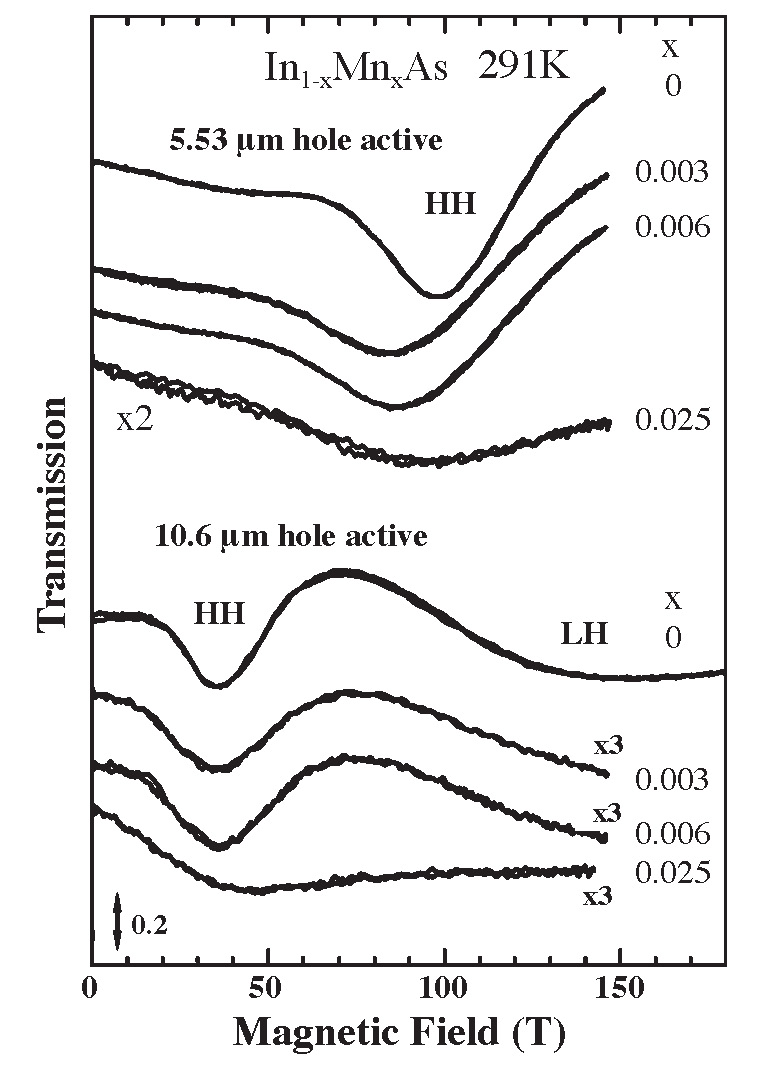}
\caption{\label{fig:ptypeInMnAsCR} Cyclotron resonance spectra for In$_{1-x}$Mn$_{x}$As with various x taken with hole-active circularly polarized light at wavelengths $5.53\mu m$ and $10.6 \mu m$\cite{InMnAs_ptype_cr}. }
\end{figure}
	
	In general it is expected that the exchange integrals $N_{o}\alpha$($\beta$) will affect the shape of the conduction (valence) band as well as the band gap.  Therefore the band mass of the carriers that reside in these bands will be sensitive to the strength of the exchange.  Since the initial cyclotron resonance measurements only determined the electron mass in n-type In$_{1-x}$Mn$_{x}$As, one might expect these measurements are only sensitive to $\alpha$. However, since InAs has a fairly small band gap, there is strong mixing between the conduction and valence band states. It is also important to note that in the $k \cdot p$ theory, the mass in a given band is proportional to the band gap (due to the amount of mixing of s and p states). To take these effects into account a theoretical model based on a $k \cdot p$ approach including the effects of  $\alpha$ and $\beta$\cite{InMnAs_cr_thry} was used to explain the data of  Fumagalli \textit{et al.}. The results of this model fit the data well, and suggest that the primary effect of the exchange is to reduce the band gap, thereby reducing the band mass. Indeed measurements of the mid-infrared absorption (see sec. \ref{sec:OptConInMnAs}) confirmed a reduction in the band gap with the introduction of Mn\cite{InMnAs_cr_thry}. Interestingly the experimental results could only be properly modeled by setting $\alpha=0.5~eV$ and $\beta=-1.0~eV$. 

\begin{figure}
\center
\includegraphics[width=18pc]{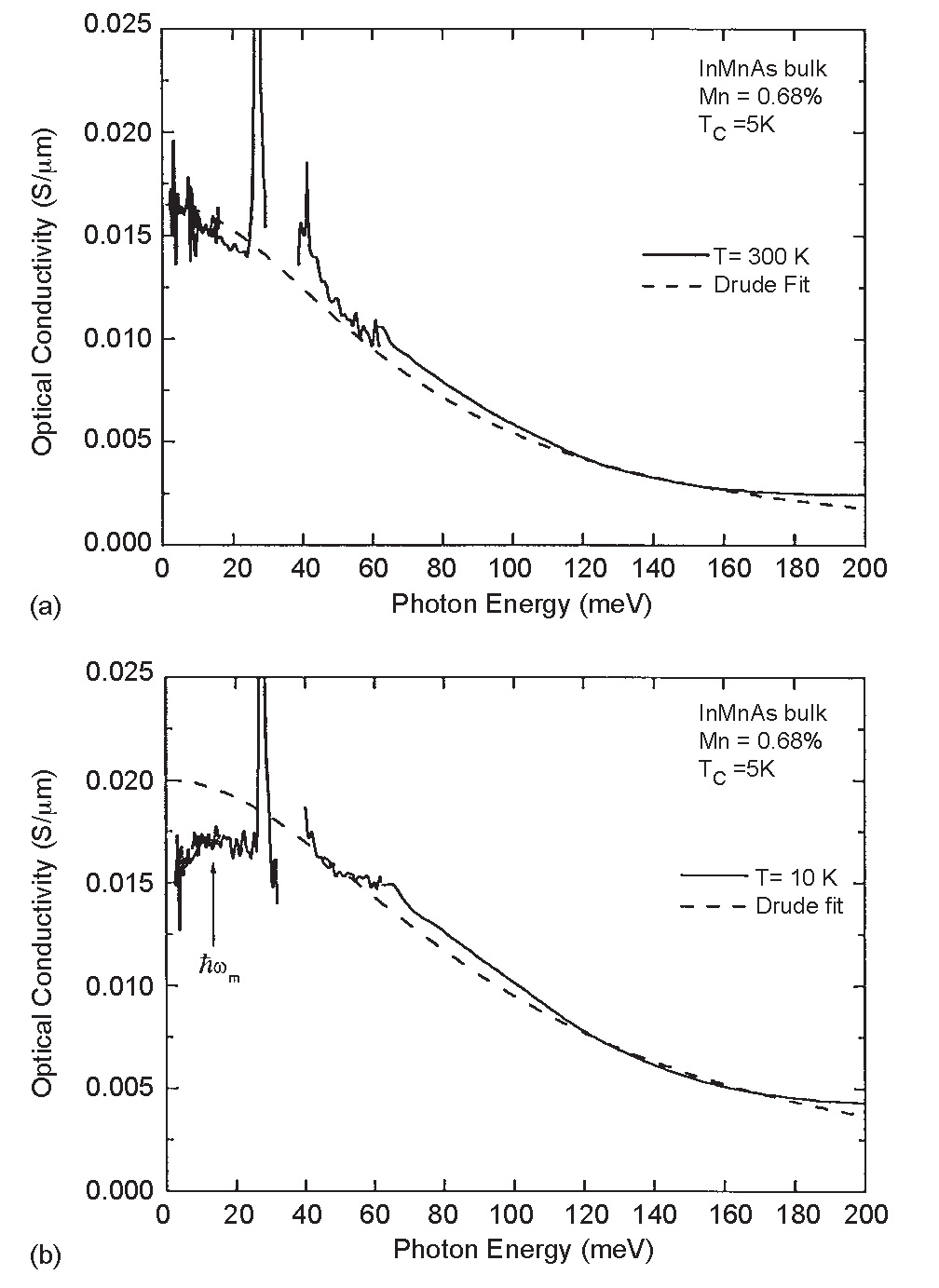}
\caption{\label{fig:InMnAsConductivity} Real part of the optical conductivity for p-type In$_{0.9932}$Mn$_{0.0068}$As. Solid lines: measured spectra with phonons producing sharp features. Top panel:  300 K  data and Drude fit (dashed lines). Bottom panel: 10 K data that displays deviations from the fit\cite{munekata_inmnAs_drude}.}
\end{figure}
	
	Given the small value of the hybridization in In$_{1-x}$Mn$_{x}$As, one might expect the holes to reside in the valance band. Experimental confirmation of this expectation by means of CR measurements of p-type In$_{1-x}$Mn$_{x}$As was exceedingly difficult since holes tend to have larger masses than electrons. Indeed, the first measurements of hole CR in In$_{1-x}$Mn$_{x}$As, shown in Fig. \ref{fig:ptypeInMnAsCR}, required fields up to 500 T, and were performed at multiple wavelengths\cite{InMnAs_ptype_cr}. These comprehensive studies revealed that the carriers have masses only consistent with conduction in the InAs valence band, and determined the cyclotron mobilities between 800 and 200 $cm^{2}/Vs$. Interestingly, unlike the n-type case, the mass did not depend on Mn doping. This also suggests that the carriers are p-type and not strongly effected by the p-d hybridization.

\subsection{Optical Conductivity of In$_{1-x}$Mn$_{x}$As}
\label{sec:OptConInMnAs}
	
	Infrared spectroscopy is another extremely powerful tool for studying the properties of III-Mn-V DMS. In these experiments one detects the intensity of the light from a spectrometer that is either transmitted through or reflected from the sample on the substrate and then normalizes it by the intensity of the light that passes through or is reflected from an equivalent substrate or passes through an open hole. From this data one can determine the absolute values of the transmission or reflection of the material in question. A unique advantage of this technique is its ability to measure the  transmission and/or reflectance of epilayers  over a broad range of frequencies at various temperatures, pressures and/or applied magnetic and electric fields. Provided the effect of the substrate is properly accounted for, one can then extract the real and imaginary parts of the complex conductivity $\hat{\sigma}(\omega)=\sigma_{1}-i\sigma_{2}$ through Kramers-Kronig analysis or by direct derivation from T($\omega$) \& R($\omega$).\cite{singley:165204}
	
	 In Fig. \ref{fig:InMnAsConductivity} the optical conductivity of ferromagnetic p-type In$_{1-x}Mn_{x}$As is shown at room temperature and at 10 K\cite{munekata_inmnAs_drude}. Interestingly at 300 K the data are described well by a simple Drude formula: $\sigma_{1}(\omega,x,T)=\frac{\Gamma_{D}^2\,{{\sigma }_{DC}}}{\Gamma_{D}^2 + {\omega }^2}$, where $\Gamma_{D}$ is the scattering rate, and $\sigma_{DC}$ is the D.C. conductivity. This is exactly what is expected in a doped semiconductor on the metallic side of the MIT in which the carriers reside in the valence band. However at low temperatures deviations from the Drude like conductivity were observed. In reference \cite{munekata_inmnAs_drude} these deviations were ascribed to the effects of localization due to the large degree of disorder and are similar to what has been seen in a conducting polymer\cite{PhysRevB.48.14884} and a disordered high temperature superconductor\cite{PhysRevB.49.12165,PhysRevLett.91.077004} The suppression of the low frequency response via strong disorder has also been reproduced theoretically\cite{PhysRevB.67.045205}. Interestingly the optical conductivity also revealed a resonant feature at $\omega\approx 1610~cm^{-1}$ (200 meV). While it has been suggested that this feature originates from metallic clusters,\cite{Noh} it also possible that the resonance results from transitions between the light (LH) and heavy hole (HH) valence bands. Since the resonance was rather weak, and has not been systematically studied across the phase diagram of In$_{1-x}$Mn$_{x}$As, its origin is still unclear. The effects of doping on an inter-band transition between the LH and HH bands will be discussed in sub-section \ref{sec:GaMnAsIR}.
	
\subsection{Time Resolved studies of In$_{1-x}$Mn$_{x}$As}
\label{TRInMnAs}
	Interest in ultra-fast (sub-picosecond) time-resolved optical spectroscopy has been growing precipitously.  An array of techniques have been employed to study relaxation dynamics in a wide range of compounds. While some of the attention to these techniques results from their potential to provide fast devices, ultra-fast studies also offer unique insight into the important excitations in various materials. Furthermore, since these measurements are often self-referencing (ie: data do not need to be normalized by measuring a separate compound with "known" properties) they can generally detect very subtle changes in the sample due to doping and/or external stimuli. Generally speaking the sample is brought out of equilibrium via an optical  femtosecond (fs) pulse whose energy is above the band gap. At a fixed time delay from the initial excitation, a second, much weaker optical pulse is used to probe the state of the sample, through a measurement of the reflection, transmission, and/or polarization rotation of the probe beam.\cite{TaylorAJ:Review,groeneveld,RevModPhys.71.S283,Awschalom_Review,KonoReview}
	
	In time-resolved studies the sample is removed from equilibrium via the creation of electron-hole pairs. These additional carriers raise the temperature of the Fermi sea via electron-electron, electron-hole, and hole-hole scattering mechanisms. This initial heating tends to occur within tens of fs, and is usually much faster than the time resolution of these experiments ($10~fs \leq \Delta t \leq 300~fs$). Consequently one generally observes a large initial change in the measured quantity. This signal then exponentially relaxes back to its initial value over a time scale (or sometimes can be fit with multiple exponentials) that reflects the transfer of heat from the Fermi sea to lattice and/or spin systems via emission of phonons and/or magnons. Thus time resolved studies provide access to the strength of the coupling between the carriers and various excitations in the system\cite{TaylorAJ:Review}. In addition, one can often manipulate the magnetic state of the sample using circularly polarized excitation that creates spin polarized carriers\cite{Awschalom_Review,KonoReview}.
	
\begin{figure}
\center
\includegraphics[width=18pc]{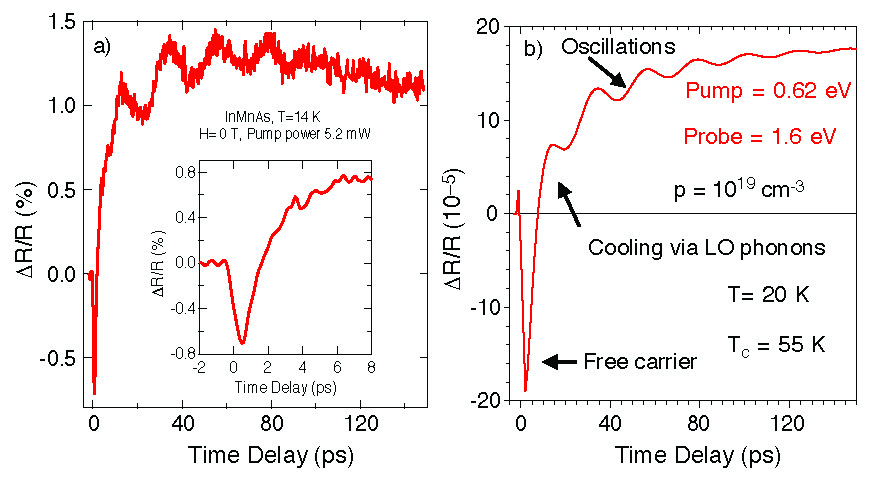}
\caption{\label{fig:InMnAsTR} Left Panel: The photo-induced change in the reflectivity as a function of time for a ferromagnetic InMnAs sample at 14 K. Right Panel: The result of a $k \cdot p$ calculation explaining the different features in the data\cite{KonoReview}.}
\end{figure}

	Time-resolved studies of In$_{1-x}$Mn$_{x}$As have revealed the importance of defects in this compound. A representative result is shown in Fig. \ref{fig:InMnAsTR}, where a ferromagnetic In$_{1-x}$Mn$_{x}$As sample has been excited with a pump at 0.62 eV well above the band-gap, and the resulting change in reflectance ($\frac{\Delta R}{R}$) has been probed with a 1.5 eV laser. Employing a pump and probe at different wavelengths provides the advantage of avoiding interference effects as well as pump-induced state filling. Interestingly Wang \textit{et al.} observed four characteristic features in the data (see Fig. \ref{fig:InMnAsTR}). The first is the sudden reduction in the reflectivity that results from the Drude absorption of the photo-generated free carriers. The second is a rise in the signal over 2 picoseconds (ps) that results in a positive $\frac{\Delta R}{R}$ signal. This relaxation channel was assigned to the trapping of charges by mid-gap states, an assertion that was confirmed by studies of low-temperature and high-temperature grown InGaAs\cite{KonoReview}. As the signal changes sign, an oscillatory signal is observed that was found to be independent of the fluence of the pump and applied magnetic field. This signal was therefore assigned to the coherent generation of acoustic modes. Finally, Wang \textit{et al.} note that the recovery occurs in hundreds of picoseconds, which they assign to the long-lived mid-gap defect states\cite{KonoReview}. 
	
\begin{figure}
\center
\includegraphics[width=18pc]{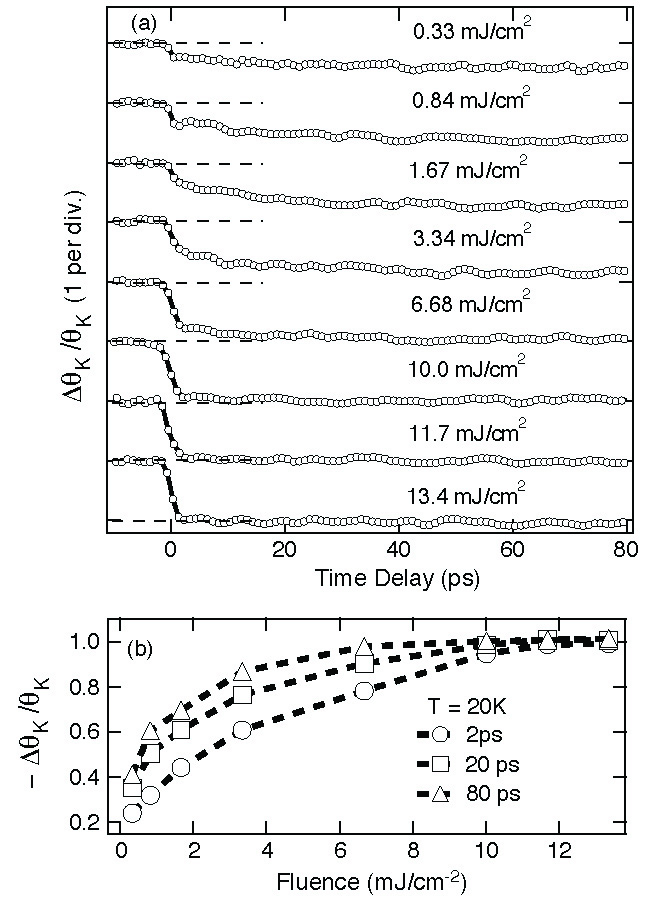}
\caption{\label{fig:InMnAsTRKerr}The change in the Kerr rotation at 1.5 eV of a ferromagnetic In$_{1-x}$Mn$_{x}$As sample after excitation at 0.62 eV. The reduction in $\Delta \Theta_{k}/\Theta_{k}$ results from demagnetization, which saturates after sufficient excitation fluence\cite{KonoReview}.}
\end{figure}

	In addition to studying changes in the reflectivity of In$_{1-x}$Mn$_{x}$As, modifications of its magnetic state have also been observed after photo-excitation. Specifically, by measuring the rotation of the polarization state upon reflection (Kerr rotation) the magnetization can be monitored as a function of time. Specifically, due to the spin-orbit coupling, the magnetization (M) results in a difference between the index for refraction ($\tilde{n}$) for opposite circular polarizations of the light. Therefore the Kerr rotation is related to changes in M and n: $\Theta\propto f(\tilde{n})M$. Again, Wang \textit{et al.} used a pump laser at 0.62 eV and detected the resulting Kerr rotation of the probe at 1.5 eV ($\frac{\Delta\Theta_{k}}{\Theta_{k}}$)\cite{KonoReview}. The $\frac{\Delta\Theta_{k}}{\Theta_{k}}$ for ferromagnetic In$_{1-x}$Mn$_{x}$As is shown in Fig. \ref{fig:InMnAsTRKerr} for various fluence values of the pump. A reduction in the Kerr rotation is observed to occur in a time less than the resolution of the experiment (220 fs), suggesting the sample magnetization is reduced after photo-excitation. This was confirmed by the observation that sign of $\frac{\Delta\Theta_{k}}{\Theta_{k}}$ reverses when the direction of the applied magnetic field is inverted. Furthermore the effect completely disappeared above $T_{C}=60~K$. To reduce any non-magnetic contributions to the Kerr rotation, the authors defined the pump-probe Kerr signal as: $\Delta\Theta_{k}=\frac{1}{2}(\Delta\Theta_{k}(-M)-\Delta\Theta_{k}(+M))$. Examining Fig. \ref{fig:InMnAsTRKerr} we also note that the change in the magnetization increases with increased pump fluence, eventually saturating at high fluence once the sample had been completely demagnetized ($\frac{\Delta \Theta_{k}}{\Theta_{k}}=-1$). In addition, the recovery time also increased greatly with additional fluence. Wang \textit{et al.} believe this provides evidence that the recovery of the ferromagnetic state occurred via the expansion of ferromagnetic domains\cite{KonoReview}. However, when the sample is completely demagnetized ferromagnetic domains must first be nucleated before the recovery can begin. It is interesting to note that despite the extensive work on the ultrafast properties of In$_{1-x}$Mn$_{x}$As, all effects observed so far in this compound appear to result from heating. Regardless, In$_{1-x}$Mn$_{x}$As could be very useful for magneto-optical storage devices. 
	
\subsection{Optical Manipulation of Magnetism in In$_{1-x}$Mn$_{x}$As}
\begin{figure}
\center
\includegraphics[width=18pc]{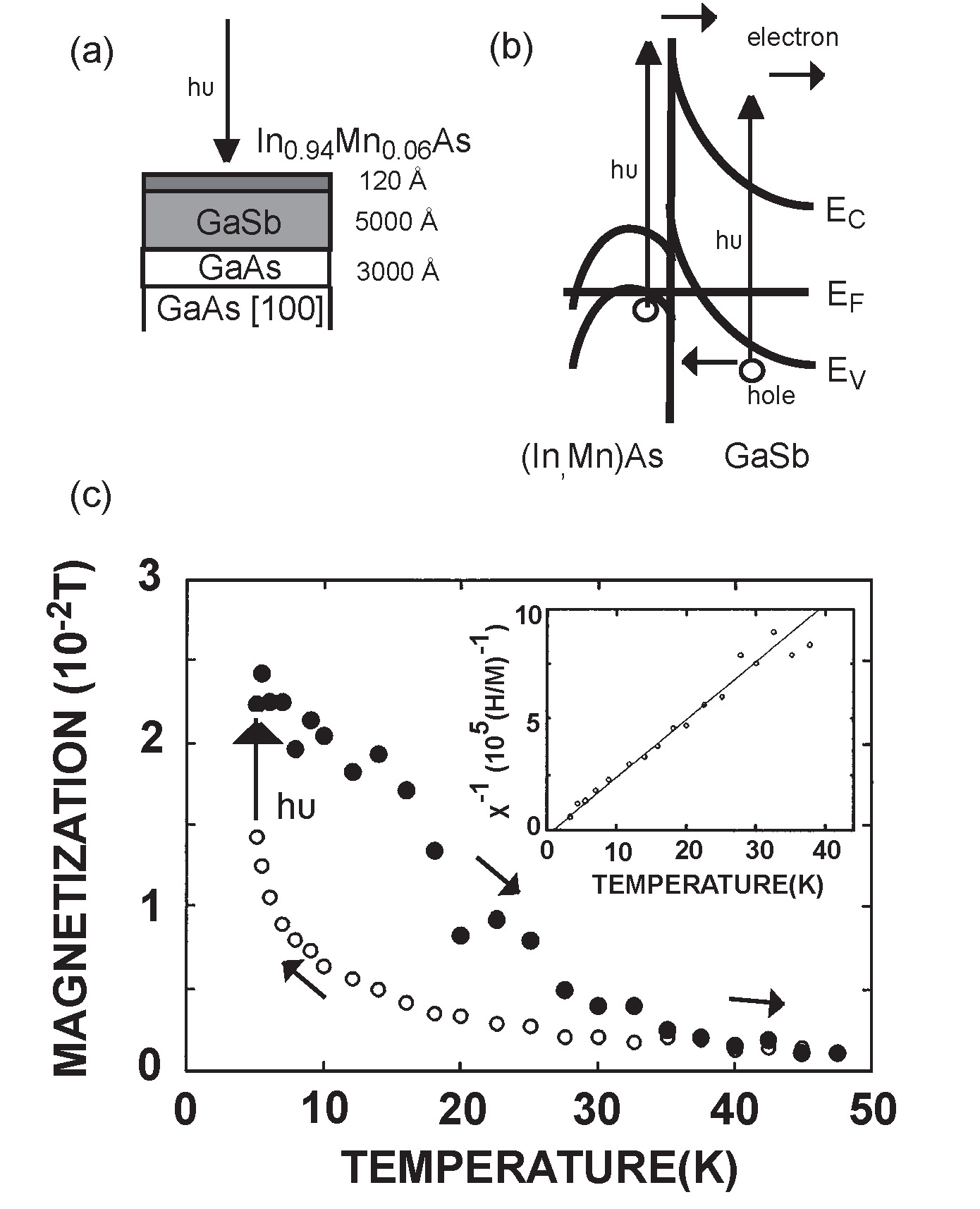}
\caption{\label{fig:munekata_photoinduced} (a) Structure of the sample with the direction of light indicated by an arrow. (b) Band edge profile of In$_{1-x}$Mn$_{x}$As/GaSb heterostructure. E$_{C}$, E$_{V}$, and E$_{F}$ denote band edges of the conduction band, valence band, and Fermi level, respectively. (c) Temperature dependence of magnetization observed during cool down in the dark (open circles) and warmup after exposure to light (solid circles) under a fixed magnetic field of 0.02 T\cite{PhysRevLett.78.4617}.}
\end{figure}

	Lastly we would like to take note of another optical experiment that clearly demonstrates the importance of holes for mediating the ferromagnetic interaction in In$_{1-x}$Mn$_{x}$As. Koshihara \textit{et al.} fabricated a lightly doped In$_{1-x}$Mn$_{x}$As film on GaSb that lies on the border of ferromagnetic order\cite{PhysRevLett.78.4617}.  Since InAs and GaSb have different energy levels with respect to vacuum and the band gap of InAs is smaller than GaSb, the In$_{1-x}$Mn$_{x}$As/GaSb structure forms a Type-II quantum well at the interface. Specifically holes are trapped in the In$_{1-x}$Mn$_{x}$As layer while electrons are trapped in the GaSb film. It was found that illuminating the samples with light whose energy was above the gap of the GaSb layer (0.8 eV), resulted in a slight increase in the conductivity and hole concentration in the In$_{1-x}$Mn$_{x}$As film. Simultaneously a large increase in the magnetization at low temperatures occured and remained long (at least 20 minutes) after the light was removed, see Fig \ref{fig:munekata_photoinduced}. Koshihara \textit{et al.} suggested these increases were long lived due to the spatial separation of the electrons (GaSb) and holes (In$_{1-x}$Mn$_{x}$As)\cite{PhysRevLett.78.4617}. Interestingly the light induced changes in the sample could all be removed by heating the film to 45 K. Other groups have also investigated this effect in ferromagnetic samples and observed either an enhancement in the magnetization\cite{LiuX:Extctd} and/or reduction in the coercive field upon illumination.\cite{oiwa:518}
	
	While these experiments raise some interesting questions, namely why is the effect removed by raising the T, it also suggests the potential to use DMS as a unique memory device. Interestingly, a similar effect was observed when additional holes were injected into the In$_{1-x}$Mn$_{x}$As layer via an electric field.\cite{spinFET1,spinFET2}

\section{Ga$_{1-x}$Mn$_{x}$As}

	Ga$_{1-x}$Mn$_{x}$As is often referred to as a ``prototypical'' ferromagnetic semiconductor. GaAs is a well-characterized semiconductor used in a variety of digital signal processing circuits, telecommunication systems, and optoelectronics. In part, this proliferation of GaAs in modern technology has driven the large body of both experimental and theoretical work over the last decade that has focused specifically on the properties of Ga$_{1-x}$Mn$_{x}$As epilayers. In order to incorporate enough Mn to make the samples ferromagnetic, Ga$_{1-x}$Mn$_{x}$As is grown at temperatures well below normal, resulting in a large density of the double donor As$_{Ga}$.\cite{myers:155203} Furthermore at high doping levels, the Mn begin to sit interstitially (Mn$_{i}$) also forming a double donor\cite{OhnoGMSFirst,OhnoH:Molbea,IyeY:Mettam,Mni1,Mni2}. In spite of these difficulties, recent advancements in growth strategies have resulted in a reduction in the total density of As$_{Ga}$ and post growth annealing has been shown to remove Mn$_{i}$, leading to enhanced carrier density and Curie temperature\cite{Mni1,Mni2,annealnot1,annealnot2,annealpenn,Mnsurface1,Mnsurface2,Mnsurface3}. Recently, a reduction of the growth temperature down to 200 C has allowed Chiba \textit{et al} to introduce as much as 20 $\%$ of Mn in 3-5 nm epilayers primarily on the substitutional sites.\cite{chiba:122503} Due to the intense efforts focused on Ga$_{1-x}$Mn$_{x}$As, it now has one of the highest Curie temperatures among all Mn-doped III-V compounds reaching 172 K in thin film samples\cite{JungwirthMaxTcGaMnAs}, and 250 K in GaAs/p-AlGaAs  $\delta$-doped heterostructures\cite{nazmul:017201}. 

	The nature of the conducting carriers mediating the ferromagnetic state in Ga$_{1-x}$Mn$_{x}$As has been called ``a central open question'' in the field\cite{zhang:125303}. Two opposing proposals of the electronic structure of metallic Ga$_{1-x}$Mn$_{x}$As include: i) a picture of mobile holes residing in nearly unperturbed valence band (VB) of the GaAs host and ii) a scenario involving persistence of the impurity band (IB) on the metallic side of the metal insulator transition with mobile holes retaining the impurity band character. Mobile holes in the VB are the cornerstone of the p-d Zener model of ferromagnetism in DMS.\cite{DietlT:Zenmdf}  A number of magnetic properties as well as trends in the evolution of the T$_C$ across the phase diagram of Ga$_{1-x}$Mn$_{x}$As appear to be consistent with this proposal.\cite{jungwirth:809,JungwirthMaxTcGaMnAs} The impurity band in turn is the starting point of alternative descriptions of the ferromagnetic state.\cite{alvarez:045202,SanvitoGaMnAsPrb,berciu:045202,MillisPRLGaMnAs,ErwinGaMnAsPRL,rader:075202}  This latter viewpoint is supported by the vast  majority of spectroscopic studies of Ga$_{1-x}$Mn$_{x}$As capable of directly probing the electronic structure. A combination of infrared conductivity measurements in Ga$_{1-x}$Mn$_{x}$As films\cite{singley:165204,Burchannealled,singley:097203} and digital ferromagnetic heterostructures\cite{burch:125340}, ellipsometric studies\cite{burch:205208}, and hot electron photo-luminescence\cite{Ploog_HPL1,Ploog_HPL2} studies are best understood within the IB picture and therefore strongly support the notion of ferromagnetism mediated by holes in the impurity band. The validity of this conclusion is further supported by tunneling\cite{tang:047201,yakunin:216806}, MCD (see Section \ref{sec:magOptGaMnAs}) and photo-emission\cite{Fujimori_PRB_GaMnAs,Okabayashi:2001lr} experiments. IR absorption investigations of annealed samples with high $T_C$\cite{Burchannealled} are inconsistent with the predictions of a model with the holes in the valence band\cite{PhysRevB.66.041202}. This ongoing debate will be discussed more fully in Sections \ref{sec:GaMnAsIR} \& \ref{sec:controversy}. 

\subsection{Broadband Optical Studies of Ga$_{1-x}$Mn$_{x}$As}
\label{sec:GaMnAsIR}

\begin{figure}
\center
\includegraphics[width=18pc]{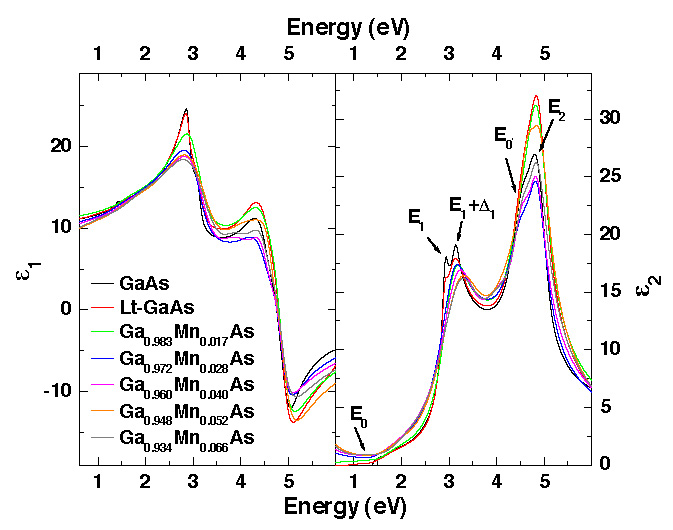}
\caption{\label{fig:elip1}The real and imaginary part of the dielectric function determined from ellipsometric measurements at 292 K for a series of Ga$_{1-x}$Mn$_{x}$As thin films with various concentrations of Mn. The plot also displays the optical constants for the GaAs substrate and an LT-GaAs film. The critical points are labeled in the conventional way, and are primarily indicative of specific k-points. A broadening of the E$_{0}$ and E$_{1}$ transitions with Mn doping is observed as well as the blue-shifting of E$_{1}$ feature (right panel).
We also note the apparent lack of change in E$_{0'}$ and E$_{2}$ transitions.
 \cite{burch:205208}}
\end{figure}

\begin{figure*}
\includegraphics[width=36pc]{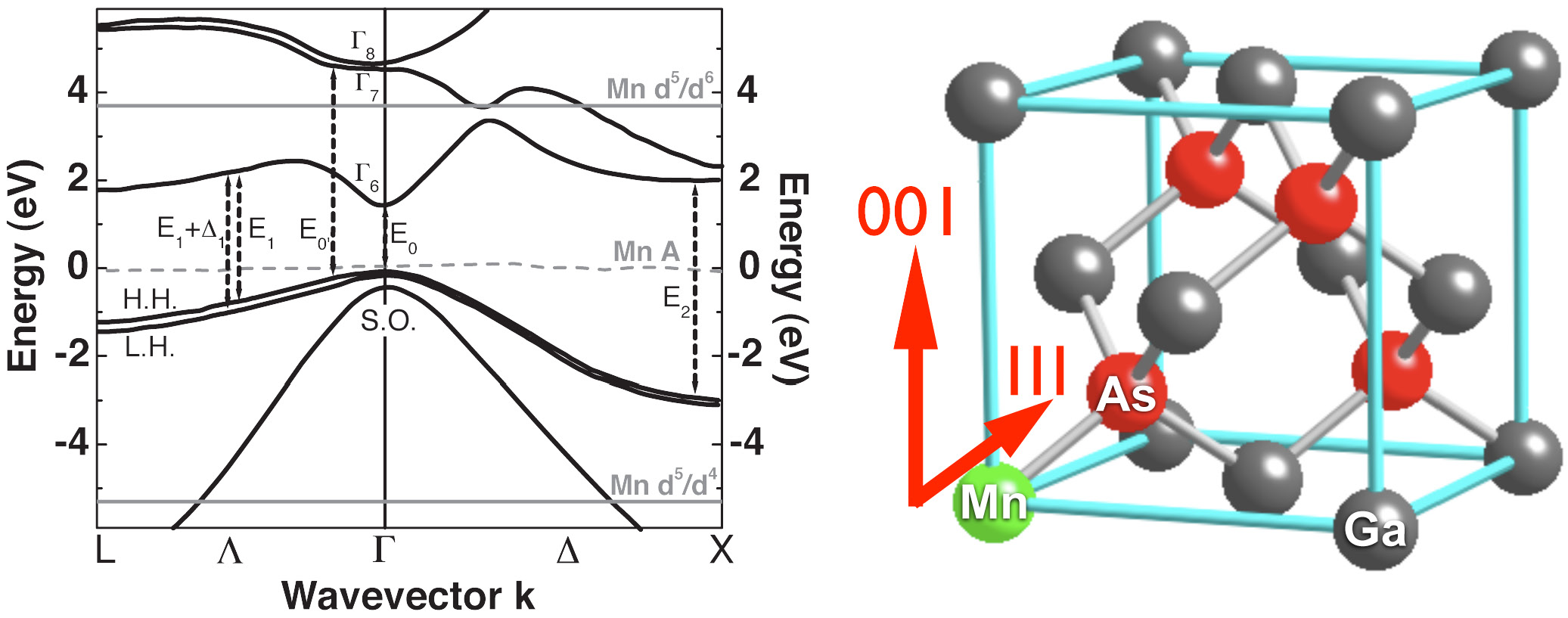}
\caption{\label{fig:bndstrc} Left: GaAs band structure and relevant critical point transitions reproduced from Ref. \cite{Cardona1}. The upper conduction bands are labeled as $\Gamma_{7,8}$ based on symmetry, while the lowest conduction band is labeled $\Gamma_{6}$. The valence bands have been labeled as H.H. for heavy-hole, L.H. for light-hole, and S.O. for split-off. Taken from Ref. \cite{rader:075202}, Mn d filled (d$^5$/d$^4$) and empty (d$^5$/d$^6$) levels are shown in grey, and the acceptor Mn A is dashed-gray. The dispersion of the Mn acceptor level is also taken from Ref. \cite{rader:075202}. The L point corresponds to the 111 direction and the X point to the 001 direction. Right: The Ga$_{1-x}$Mn$_{x}$As unit cell with the important symmetry directions labeled.}
\end{figure*}

	Spectroscopic ellipsometry experiments were pivotal in establishing the electronic structure of standard semiconductors\cite{CardonaBook}. Spectroscopic ellipsometry is particularly advantageous for the studies of materials because the optical constants can be measured directly without recourse to Kramers-Kronig analysis, and one can often obtain information on Oxide content and epilayer thicknesses. Ellipsometric studies of Ga$_{1-x}$Mn$_{x}$As focused on the transformations upon Mn doping of the dielectric constants ($\hat{\epsilon}(\omega)=\epsilon_{1}(\omega)+i\epsilon_{2}(\omega)$) above the band gap of the GaAs host\cite{burch:205208}. A brief inspection of Fig \ref{fig:elip1} shows unmistakable similarity in the gross features of the ellipsometric data between pristine GaAs and heavily doped epilayers with Mn concentrations exceeding $7\%$. The characteristic structure in the spectra of $\epsilon_1(\omega)$ and $\epsilon_2(\omega)$ are produced by inter-band transitions associated with van Hove singularities in the joint density of states labeled in Fig. \ref{fig:elip1} using a standard notation\cite{CardonaBook}. The band structure of Ga$_{1-x}$Mn$_{x}$As, including these transitions and important symetry directions are shown in Fig. \ref{fig:bndstrc}. Importantly, $E_1$, $E_{0^{\prime}}$ and $E_2$ features of the GaAs host can all be recognized in the ellipsometry data for heavily doped Ga$_{1-x}$Mn$_{x}$As samples. This result attests to the quality of the Ga$_{1-x}$Mn$_{x}$As samples. Indeed, the above characteristic features are known to be wiped out in disordered semiconductors\cite{erman:2664}. 

	Burch \textit{et al.} quantitatively studied the evolution of the critical points with doping via an analysis of the second derivative spectra\cite{burch:205208}. This approach has the advantage of amplifying the subtle features of the optical constants upon taking their derivatives. A significant blue-shifting of the E$_{1}$ critical point was observed, while the position of all other critical points were unaffected by the presence of Mn. The blue-shifting of E$_{1}$ was attributed to the hybridization of the Mn impurity band with the GaAs valence band\cite{burch:205208}.  Since the E$_{1}$ critical point originates from transitions near the L-point (ie: in the 111 direction), the authors suggested that $V_{pd}$ is k-dependent, which was recently confirmed by a combined experimental and theoretical study of the electronic structure around a local Mn atom\cite{Kitchen:2006lr}. Since the Mn-As bond is along the 111 crystallographic direction (see Fig \ref{fig:bndstrc}), $V_{pd}$ should be strongest near the L-point\cite{fiete:045212,burch:205208}. 

	It should be noted that Zhang and Das Sarma have recently shown that if Fermi energy resided in the valence band, then a significant red-shifting of the fundamental band gap should be observed. Specifically, the Coulomb interaction between the holes would lead to a re-normilization of the critical points.\cite{zhang:125303} However, in Mn-doped samples the fundamental band gap of GaAs ($E_0$) was not detected in the ellipsometry experiments, though this is not due to the Mn doping per se. In fact the $E_0$ structure is not seen is GaAs without any Mn when it is grown at low temperature. 
	
	The ellipsometry results presented in \cite{burch:205208} provide additional insights into the broadening of the band gap of GaAs grown at low temperatures. In previous studies of these samples it was determined that the broadening was, in part, the result of transitions either beginning (in the case of n-type LT-GaAs) or ending (in the case of p-type Ga$_{1-x}$Mn$_{x}$As) in the As$_{Ga}$ impurity states.\cite{singley:165204} Additional information provided by the $\hat{\epsilon}(E>1.5eV)$, suggested that the broadening of the fundamental gap also the results from a relaxation of the requirement of momentum conservation. This relaxation is due to the presence of impurities that provide additional scattering mechanisms. Since transitions are no longer required to be direct, states in the valence band that are not at the zone center can contribute to transitions which end at the zone center. Ultimately this results in a broadening of transitions and a transfer of spectral weight from higher energies to lower ones, as is seen in Fig. \ref{fig:elip1}. We note that a similar result is found in GaAs damaged by Ion-implantation.\cite{Aspnes1} Nonetheless, as discussed in Section \ref{sec:magOptGaMnAs}, MCD studies suggest that the fundamental band gap of Ga$_{1-x}$Mn$_{x}$As only slightly blue-shifts with doping  in contradiction to predictions of the valence band scenario. 

\begin{figure}
\center
\includegraphics[width=18pc]{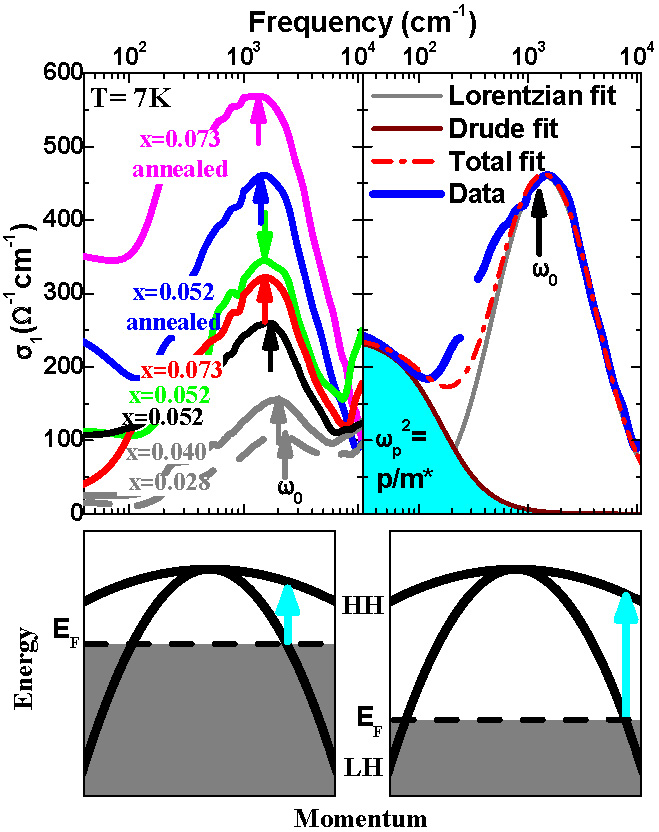}
\caption{\label{fig:IrAnneal} (Bottom Panels) Diagrams of the electronic structure of Ga$_{1-x}$ Mn$_x$As assuming E$_{F}$ lies in the light (LH) and heavy hole (HH) bands. (Left-Bottom) at low x, a transition occurs between the HH and LH bands. (Right-Bottom) as the carrier density is increased, E$_{F}$ moves farther into the LH and HH bands, causing the transition to blue-shift. Top panels: the real part of the conductivity for a series of Ga$_{1-x}$Mn$_{x}$As samples studied. An increase in $\sigma_1(\omega)$ occurs with larger Mn concentration x and/or annealing. Top-right panel shows the results of fitting $\sigma_1(\omega)$ for the annealed, $x=0.052$ sample using a Drude-Lorentz model\cite{Burchannealled}.}
\end{figure}

	We now turn our attention to studies of the optical constants below the fundamental band gap of the GaAs host. As pointed out above, Mn doping leads to the formation of a metallic state in Ga$_{1-x}$Mn$_{x}$As films\cite{OhnoGMSFirst}. A counterpart infrared effect is a dramatic change of the optical conductivity throughout the entire frequency range within the band gap of the GaAs host. These changes were first investigated by Nagai \textit{et al.} who examined the absorption coefficient\cite{NagaiY:Spipdf}. This study was later followed by an analysis of the {\it optical conductivity} of Ga$_{1-x}$Mn$_{x}$As by Singley \textit{et al.}\cite{singley:165204,singley:097203}. Most recently, a comprehensive study of both as grown and annealed films has been reported by Burch \textit{et al.} (see Fig \ref{fig:IrAnneal})\cite{Burchannealled}. With the exception of one work, all Ga$_{1-x}$Mn$_{x}$As  samples investigated so far using infrared methods were prepared using low temperature molecular beam epitaxy (MBE). The only exception is the work of Seo \textit{et al.} in films obtained using implantation of Mn ions into a GaAs film\cite{seo:8172}.
	
	Ferromagnetic films of Ga$_{1-x}$Mn$_{x}$As grown by MBE reveal two new features in the intra-gap conductivity. The first is a broad resonance initially centered at approximately 2000 cm$^{-1}$, whose center energy is doping dependent. The second key feature is the presence of finite conductivity in the limit of $\omega\rightarrow 0$, signaling metallic behavior. The oscillator strength of both features increases with additional Mn doping. The primary impact of annealing is to further increase the spectral weight of these two features while their frequency dependence remains mostly unchanged. Due to the reduction in defect concentration, annealing results in a slope in the conductivity for $\omega<200~cm^{-1}$ as prescribed by the simple Drude model.\cite{Burchannealled}    

	An observation of a resonance in the mid infrared is consistent with either of the two fundamentally different views of the electronic structure of Ga$_{1-x}$Mn$_{x}$As. Specifically assigning mobile holes at the Fermi energy ($E_F$) to the impurity band (IB) or to the valence band (VB) of the GaAs host will produce a mid infrared (MIR) feature.  Indeed, if the holes reside in the IB the MIR resonance results from transitions between the IB and VB\cite{mahadevan:115211,alvarez:045202,SanvitoGaMnAsPrb,majidi:115205,berciu:045202,MillisPRLGaMnAs,DasSarmaOptCond}. The 2000 cm$^{-1}$ mode clearly seen in the top-left panel of Fig \ref{fig:IrAnneal} occurs close to Lyman transitions produced by dilute concentrations of Mn in GaAs single crystals\cite{PhysRevLett.18.443}. However, if $E_F$ lied in the valence band of the GaAs host, then one expects a MIR resonance to result from transitions between the light and heavy hole bands. This latter interpretation was originally proposed by Nagai \textit{et al}\cite{NagaiY:Spipdf} and was followed up by calculations including the role of disorder and a more realistic band structure \cite{PhysRevB.67.045205,PhysRevB.66.041202,aguado:195201}. In fact, calculations of the optical conductivity carried out within the VB framework reproduce the shape of the experimental data. Nevertheless one can discriminate between the two scenarios by monitoring the doping dependence of the position of the MIR peak.
	
	As illustrated in the bottom panels of Fig \ref{fig:IrAnneal} the VB picture implies hardening of the resonance with increasing doping. This natural conclusion is supported by the calculations of Sinova \textit{et al.} within the VB scenario\cite{PhysRevB.66.041202}. However, data reported by Burch \textit{et al.} show a pronounced softening of the MIR resonance that is only consistent with the IB model (see Fig. \ref{fig:IBGraph})\cite{Burchannealled}. Further support for the impurity band scenario originates from an analysis of the optical effective masses associated with the free carrier absorption. These masses inferred from the oscillator strength sum rule yield m$^{*}$ on the order of 10 times $m_{e}$\cite{singley:165204,Burchannealled,singley:097203}, a result recently confirmed by studies of the mobility of Ga$_{1-x-y}$Mn$_{x}$AsBe$_{y}$ and Ga$_{1-x}$Mn$_{x}$As$_{1-y}$P$_{y}$\cite{alberi:075201}. These high masses should be contrasted with the VB mass ($\sim0.38~m_{e}$ \cite{songprakob:171}).  Thus, metallic Ga$_{1-x}$Mn$_{x}$As epilayers reveal the behavior in stark contrast to the expectations of the valence band scenario. 

Jungwirth et al.\cite{jungwirth:PRB} have proposed that the redshift of the mid-IR resonance in Fig. \ref{fig:IBGraph} can be anticipated as a consequence of increased screening in samples with enhanced metalicity. The authors find support for this statement with finite-size exact-diagonalization calculations originally reported in Ref.\cite{PhysRevB.67.045205}. These calculations indeed show that minute red shift may occur in highly disordered samples. A much more pronounced effect in these calculations is a dramatic narrowing of the mid-IR resonance that is required to reproduce barely noticeable red shift. This is not consistent with experiments revealing redshift without any significant modifications of the lineform. Furthermore, data do not uncover any apparent empirical relationship between the position of the  
resonance and $\sigma_{DC}$ that are anticipated within the framework of a proposal by Jungwirth et al.\cite{jungwirth:PRB} On these grounds, we conclude that the doping trends seen in the broad-band studies of electromagnetic response of Ga$_{1-x}$Mn$_{x}$As cannot be reconciled with the calculations based on the valence band scenario. 

It is worth pointing out that only the valence band scenario of the electronic response of Ga$_{1-x}$Mn$_{x}$As has reached the sophistication required for detailed comparison with existing optical experiments. Considerably less effort has been dedicated so far to the theoretical exploration of the properties implied by the impurity band picture. It remains to be seen if this latter picture can sustain similar experimental scrutiny going beyond qualitative discussion of properties implied by the impurity band. As we will argue below a realistic picture of electronic and magnetic properties of Ga$_{1-x}$Mn$_{x}$As must account for coexistence of extended and localized properties of Mn-induced holes: a task that is beyond the existing theoretical approaches. We will return to the discussion of these issues in Subsection \ref{sec:controversy}. 

	Ferromagnetism at high transition temperatures in Ga$_{1-x}$Mn$_{x}$As occurs at extremely high Mn doping levels close to the boundary where Mn begins to form clusters in the GaAs host. It is therefore important to address the issue of whether the optical properties discussed above represent the intrinsic electromagnetic response of Ga$_{1-x}$Mn$_{x}$As with Mn primarily entering substitutionally in the GaAs lattice.  This issue is particularly important since some of the characteristic absorption features seen in Fig \ref{fig:IrAnneal} are consistent with the suggestion of amorphous MnAs clusters imbedded in low temperature grown GaAs (LT-GaAs). It has been reported that implantation of Mn ions in LT-GaAs films result in MnAs clusters that produce a resonant structure reminiscent of the optical conductivity of Ga$_{1-x}$Mn$_{x}$As films grown by molecular beam epitaxy\cite{seo:8172}. However in the implanted films the resonance is centered around 0.8 eV, and therefore appears to be inconsistent with what is found in the films grown by molecular beam epitaxy. Finally we note that if the feature seen in the MIR were due to metallic clusters inside the GaAs matrix, the resonance would be centered at $\omega_{p}/\sqrt{2\epsilon^{h}_{1}+1}$ where $\omega_{p}$ is the plasma frequency of the metallic cluster,\cite{seo:8172} which is proportional to the carrier density divided by their mass, and $\epsilon^{h}_{1}$ is the dielectric constant of the host. Therefore the red-shifting with doping would result from one of two unlikely scenarios: a decrease in $\omega_{p}$ of the metallic spheres or an increase in the dielectric constant. 
	
\subsection{Light Scattering Studies of Ga$_{1-x}$Mn$_{x}$As}

	Inelastic light scattering has proven to be a useful technique for studying bosonic modes in a broad range of materials. In general a laser polarized along a specific crystallographic direction is reflected off the surface of a sample. The incoming photons undergo a scattering event wherein they either absorb (Stokes) or emit (Anti-Stokes) a boson in the sample, resulting in an energy shift of the reflected photon. By measuring a specific polarization of the reflected light, one can select out different modes via the optical selection rules. This technique, often called Raman scattering, has measured the charge state of the Mn ion as a function of x in Ga$_{1-x}$Mn$_{x}$As. 
	
\begin{figure}
\center
\includegraphics[width=18pc]{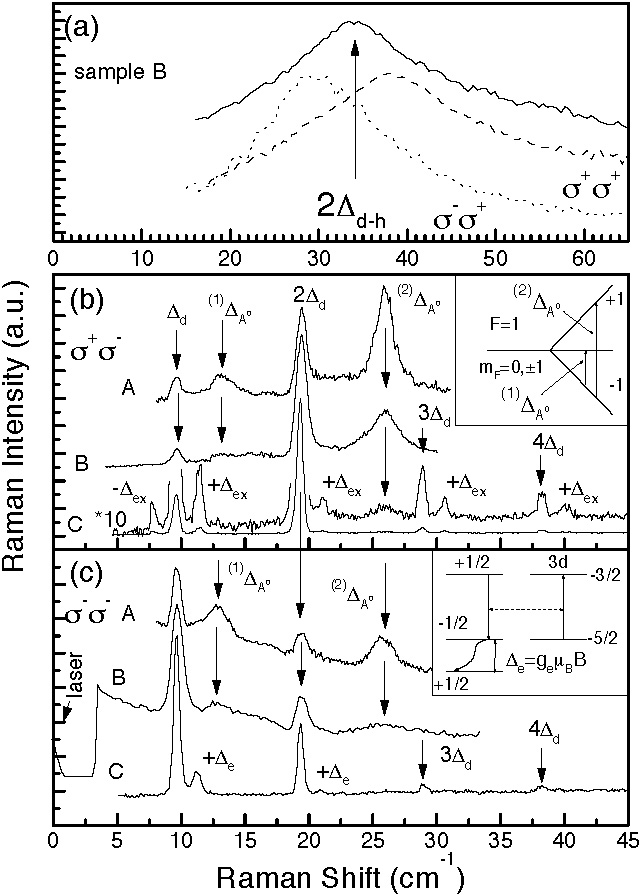}
\caption{\label{fig:sapega_raman} Raman spectra of Ga$_{1-x}$Mn$_{x}$As samples A ($N_{Mn}=6\times10^{17}~cm^{-3}$), B ($N_{Mn}=8\times10^{17}~cm^{-3}$) and C ($N_{Mn}=3.5\times10^{18}~cm^{-3}$) measured in resonance with the acceptor-exciton energy level. (a) The 2$\Delta_{d-h}$ feature in sample B is related to the transition between the $F=1$ and $F=2$ acceptor states at 0 T (solid line) and 8 T. (b) and (c) Stokes Raman spectra at 10 T for two circular polarization geometries, labels are explained in the text. The inset in (b) shows the splitting of the F=1 acceptor ground state into Zeeman sub-levels. The inset in (c) illustrates the origin of the satellite resonance labeled $\Delta_{e}$ in a two-step Raman process\cite{SapegaVF:SpiRsM}.}
\end{figure}

	The first such study was performed on paramagnetic Ga$_{1-x}$Mn$_{x}$As single crystals grown by the Czochralski technique, with Mn concentrations of 0.6-3.5 $\times10^{18}~cm^{-3}$ \cite{SapegaVF:SpiRsM}. The authors focused on Spin-flip Raman scattering (SFRS) wherein the spin of either the total angular momentum of the hole and Mn complex or just that of the d-shell itself is changed. One expects the ground state of the total angular momentum of the Mn-hole complex to be $F=|S_{d}-J_{h}|=1$, where $S_{d}=5/2$ is the total angular momentum of the Mn d-shell and $J_{h}=3/2$ for the holes. This should lead to features in the Raman spectra due to transitions from the $|F=1,m_{F}=-1>$ level to the $|F=2,m_{F}=-2>$ and $|F=2,m_{F}=-1>$ states. In Fig \ref{fig:sapega_raman}(a) a typical Raman spectra is shown with a feature due to a change in the angular momentum of the Mn-hole complex from F = 1 to F = 2 at 2$\Delta_{d-h}$. It should be noted that the energy of this resonance depends on the p-d exchange strength. 

	Interestingly by using different polarization configurations in magnetic field additional features were observed in the Raman spectra (see Fig \ref{fig:sapega_raman}(b)). The center energy of these new resonances shift linearly in field and tend to zero as the applied field is removed. Spin flips of the acceptor that conserve the total angular momentum are observed as two broad features labeled $^{(1,2)}\Delta_{A^{0}}$ and shift with applied magnetic field as $(1,2)\times g_{A_{0}}\mu_{B}B$, where $g_{A_{0}}=2.74$ is the g-factor of the acceptor, and $\mu_{B}$ the Bohr magneton. The difference in their energy is related to whether the flip is between states of $\Delta m_{F}=\pm1$ or $\Delta m_{F}=\pm2$. In the samples with heavier doping levels these features disapear due to the formation of the impurity band. However the sharp features in the spectra labeled $n\Delta_{d}$, with n=1,2..., remain in samples with larger Mn content and gain satellite features seen on the Stokes side (labeled $+\Delta_{e}$ and $\Delta_{ex}$) and Anti-Stokes side (at $- \Delta_{ex}$) with larger Mn content. These resonances result from transitions within the Mn d-shell due to its interactions with the excitons ($\Delta_{ex}$) and photo-excited electrons ($+\Delta_{e}$) created by the laser. Therefore one observes the flipping of the Mn spins $\Delta_{d}$ as well as the exciton ($\Delta_{ex}$) and its electron ($\Delta_{e}$). Furthermore the number of $\Delta_{d}$ resonances depend on the number of Mn ions seen by the exciton. The field dependence of $\Delta_{e}$ and $\Delta_{ex}$ is characteristic of excitons experiencing the external and exchange fields, and indicate the Mn ion is in a $d^{5}$ state. 

	A later Raman study of MBE grown films focused on the effects of increasing the Mn density. The 2$\Delta_{d-h}$ resonance was only observed in samples with lower doping levels and its intensity was significantly temperature dependent, which results from thermal activation of the holes from the $F=1$ to $F=2$ states and/or to the valence band \cite{PhysRevB.66.075217}. In addition, the intra-d shell resonance was observed in the samples with small Mn levels, but only above 50 K, and its intensity was found to drop above 150 K. This temperature dependence is likely the result of the holes being tightly bound to the Mn at low temperatures, resulting in spin flips of the Mn-hole complex only. At higher temperatures the holes are thermally activated and spin-flips of the Mn ion emerge. Interestingly in the samples with heavier doping levels of Mn, the 2$\Delta_{d-h}$ resonances disappeared and a new feature appeared at $\Delta_{2}$. This $\Delta_{2}$ feature was only observed for $T>90~K$. From its temperature and polarization dependence, Sapega \textit{et al.} conclude $\Delta_{2}$ is a transition within the Mn $^5T_{2}$ levels that are split by the dynamic Jahn-Teller effect\cite{PhysRevB.66.075217}. Such a transition is only possible if the Mn is in a d$^4$ configuration, and therefore these studies suggest that with increasing concentration, the Mn goes from being in a d$^5$ to d$^4$ state. Furthermore Sapega \textit{et al.} concluded that the holes begin in acceptor levels with angular momentum $J_{h}=3/2$ but broaden into an impurity band. 

Two recent Raman studies of Ga$_{1-x}$Mn$_{x}$As have taken an alternate approach to the spectra\cite{Samarth_Raman,PhysRevB.66.205209}. They focused on the Raman spectra near the longitudinal (LO) and transverse (TO) optically active phonon modes. Both observed a mode due to a coupling between the LO phonon and free carrier plasmons, from which they can extract the carrier densities. This technique is quite powerful as it can determine the carrier density accurately without needing to apply large magnetic fields, as is required in Hall measurements. Interestingly, one of these studies also observed a shifting in the LO and TO modes, which they attributed to the alloying effect of Mn\cite{PhysRevB.66.205209}.  

\subsection{Magneto-Optical Studies of Ga$_{1-x}$Mn$_{x}$As}
\label{sec:magOptGaMnAs}
	Insights into the exchange mechanism between the holes and local moments in DMS have been supplied by studies of their magneto-optical properties. Magneto-circular dichroism (MCD) measurements provide access to the Zeeman splitting of the conduction and valence bands, such that the strength and nature of the exchange is determined\cite{furdyna_review}. In addition, magneto-optical techniques have been used to examine the domain structure and its evolution upon magnetization reversal in Ga$_{1-x}$Mn$_{x}$As.\cite{PhysRevLett.90.167206,thevenard:195331} Furthermore, it has been shown magneto-optics are also quite powerful for studying the in-plane magnetization dynamics. Indeed, magneto-linear dichroism where the difference between the absorption or reflection of the light polarized parallel and perpendicular to the magnetization direction delivers direct access to the size and direction of the in-plane magnetization component.\cite{moore:4530,kimel:227203} 
	
	Perhaps the most powerful aspect of magneto-optical experiments lies in their ability to discriminate which electronic states strongly hybridize the local moments.\cite{furdyna_review,KojiAndo06302006} Specifically, an MCD signal at a given energy indicates a difference between the spin up and spin down bands involved in the optical transition, which results from their hybridization with the local moments. This is particularly useful since an MCD signal at the fundamental gap of the host semiconductor provides a highly reliable measure of intrinsic ferromagnetism. Indeed, if clusters of another phase exist within the sample, they will be detected by measurements of the magnetic moment of the sample, but they will not result in a spin-splitting of the host material bands. This is due to the fact that the moments of the secondary phase do not hybridize with the host states. Therefore performing MCD measurements in addition to standard magnetization measurements is critical for demonstrating a material is truly a diluted magnetic semiconductor and not simply a two-phase compound\cite{AwschalomMCD,KojiAndo06302006}. As discussed below, by measuring the MCD signal, one can quantify the strength of the hybridization between the states mediating the ferromagnetism and the local moments, as well as determine wether it is ferromagnetic or antiferromagnetic. 
	
	Magneto-optical experiments are generally performed in a Faraday configuration (light propagates parallel to the external magnetic field and perpendicular to the sample). One then measures either the difference in the absorption or reflection of the left and right circularly polarized light. The earliest magneto-optical studies of Ga$_{1-x}$Mn$_{x}$As focused on the MCD spectra, which we define as $MCD=\frac{\alpha^{-}-\alpha^{+}}{\alpha^{-}+\alpha^{+}}$, where $\alpha^{+(-)}$ are the absorptions for right(left) circularly polarized light. It should be noted that many of these studies were conducted in reflection and defined MCD as $\frac{R^{-}-R^{+}}{R^{-}+R^{+}}$, where $R^{+(-)}$ are the reflection coefficients for  right(left) circularly polarized light. The MCD spectra are quite useful as they can be related  to the s-d and p-d exchange constants, for instance in reflection\cite{AndoDerivMCD}: 
	\begin{equation}
	\label{eq:MCD}MCD\propto -\frac{g_{exc}\mu_{B}B+N_{0}(\alpha-\beta) S_{Z}(T,B)}{R}\frac{dR}{d\omega}
	\end{equation}
	where $g_{exc}$ is the excitonic g-factor, and  $S_{Z}(T,B)$ is the average value of the Mn spin per Mn ion, R is the reflectance and $\omega$ is the frequency at which the measurement is made. Therefore the MCD spectra provides a measurement of both the strength and nature (ie: antiferromagnetic or ferromagnetic) of the exchange between local moments and the electrons/holes in the system. However it should be noted that this approach assumes a single component in the spectra and relies on a knowledge of the frequency dependence of the reflectance/absorption. 

\begin{figure}
\center
\includegraphics{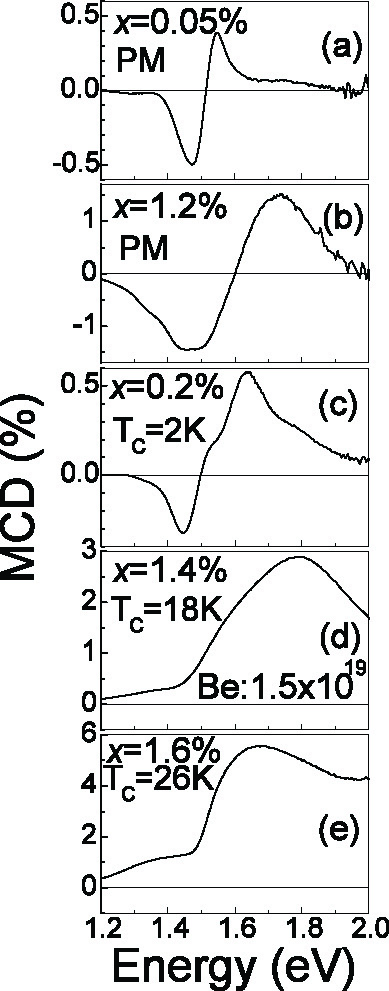}
\caption{\label{fig:ChakarvortyMCD} Evolution of the MCD spectra taken at T=1.8 K and in a magnetic field B=5.0 T. Samples with x=0.0005 and 0.012 are paramagnetic and
samples with x=0.002, 0.014, and 0.016 are ferromagnetic, with $T_{C}$s of 2 K, 18 K, and 26 K respectively. The x=0.012 sample was grown at lower
temperature than the other samples, resulting in a much higher compensation
level. The horizontal line represents the zero level for
the MCD signal. All five layers had nominally the same thicknesses of 0.3$\mu$m.\cite{chakarvorty:171118}.}
\end{figure}
	
	Due to the importance of the value of the exchange splitting to models of the ferromagnetism in DMS, the MCD spectra of Ga$_{1-x}$Mn$_{x}$As have been extensively studied as a function of doping and/or temperature.\cite{SzczytkoJ:spdeiG,AndoK:Magetf,OhnoMCD,hartmann:233201,furdynaMCD,AwschalomMCD,lang:024430,chakarvorty:171118,ando-2007,ando:067204} Initially this vast array of experiments produced a range of exchange values $-1.0 \leq N_{0}\beta \leq 2.5$. These initial studies lead some to believe the value of the exchange changed sign with increased Mn concentrations and/or carrier densities. Furthermore, the MCD signal seen in ferromagnetic Ga$_{1-x}$Mn$_{x}$As samples was quite surprising since it was positive, suggesting that the p-d exchange is ferromagnetic, rather than antiferromagnetic as in all II-Mn-VI compounds.\cite{SzczytkoJ:spdeiG,furdyna_review} Numerous explanations were put forward to explain the change in the sign of the MCD as well as the apparent ferromagnetic exchange.\cite{SzczytkoJ:spdeiG,OhnoMCD,PhysRevB.67.115203,PhysRevB.64.075306,furdynaMCD,hartmann:233201} Nonetheless, it should be noted that the original studies spanned a large range of carrier and Mn densities. Furthermore the derivation of the exchange constants as well as the explanations of its sign and doping dependence generally assumed the MCD spectra resulted from a single optical transition (ie: only between two bands).  
	
\begin{figure}
\center
\includegraphics[width=18pc]{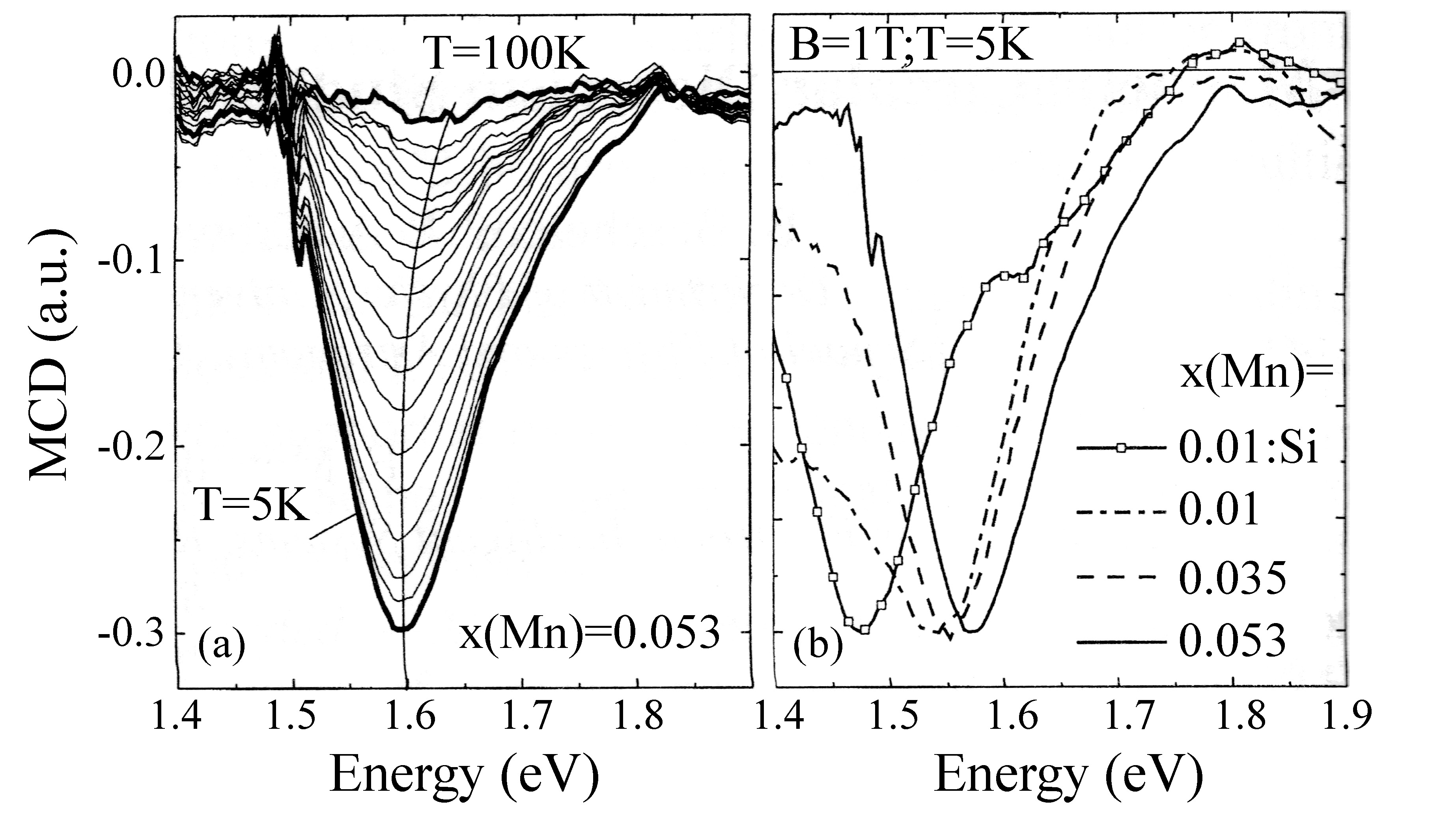}
\caption{\label{fig:AwschalomMCD} MCD spectra for Ga$_{1-x}$Mn$_{x}$As, all curves are obtained by dividing by their value at 1.85 eV, then subtracting the data at $T>T_{C}$. (a) T dependence of the x=0.053 sample.  (b) x dependence at 5 K and 1 T \cite{AwschalomMCD}.}
\end{figure}
	
	The origin of this apparent confusion has recently been clarified by the systematic MCD data studies of Chakarvorty \textit{et al.}\cite{chakarvorty:171118} and Ando \textit{et al.}\cite{ando-2007,ando:067204}. Representative results are presented in Fig \ref{fig:ChakarvortyMCD}, where the carrier density increases as one moves from the top to bottom panels. The spectra in the top panel for a very small doping level (x=0.0005) correspond well with that observed in GaAs, namely a sharp feature near the fundamental gap that changes sign as a function of energy. However upon increasing the Mn and/or carrier density, the spectra appear to broaden and reveal a broad positive MCD signal. 

\begin{figure}
\center
\includegraphics[width=18pc]{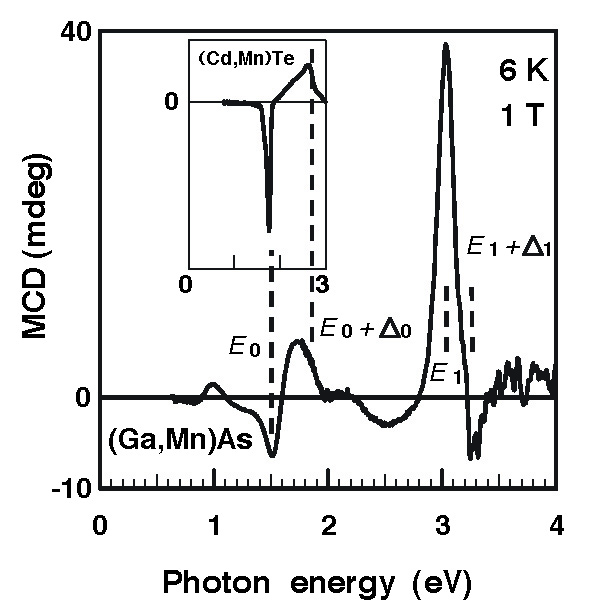}
\caption{\label{fig:AndoMCD} MCD spectrum of paramagnetic (Main Panel) Ga$_{0.996}$Mn$_{0.004} $ at 6 K and Cd$_{0.92}$Mn$_{0.08}$Te at 15 K and 1 T. Critical points for GaAs and CdTe are labeled. The horizontal scale of the inset has been adjusted, such that the E$_{0}$ and E$_{0}+\Delta_{0}$ critical points overlap in the two materials\cite{ando-2007,ando:067204}.}
\end{figure}

	Two studies of the MCD spectra of Ga$_{1-x}$Mn$_{x}$As have pointed out that the data contain two features with opposite sign and very different shapes.\cite{AwschalomMCD,ando-2007,ando:067204} Namely a broad positive MCD signal is observed around 1.9 eV, the amplitude of which scales with the magnetization and doping level. This is not surprising in light of equation \ref{eq:MCD}, which suggests the MCD signal should be proportional to the magnetization times a temperature, frequency and magnetic field independent term (ie: $MCD\propto -N_{0}(\alpha-\beta)~\times<S_{Z}(T,B)>$). Furthermore, Beschooten \textit{et al.} showed one could remove the broad background by collapsing the spectra. Specifically they divided the MCD results by their value at 1.85 eV, upon which a sharp feature emerged near the fundamental band gap of GaAs (E$_{0}$ critical point). This resonance is shown as function of temperature for the $x=0.053$ sample in the left panel of Fig \ref{fig:AwschalomMCD}, where high temperature normalized MCD spectra has been subtracted from the scaled MCD. Interestingly this additional resonance only emerges below $T_{C}$ and has the opposite sign as the 1.9 eV broad feature. Ando \textit{et al.} have recently pointed out that the polarity and shape of the MCD spectra near the E$_{0}$ and E$_{0}+\Delta_{0}$ critical points in dilute Ga$_{1-x}$Mn$_{x}$As samples closely resembles that of Cd$_{1-x}$Mn$_{x}$Te (see Fig. \ref{fig:AndoMCD}). Furthermore, this negative MCD signal from the fundamental band gap appears to remain to the highest doping levels and slightly blue shift (see right panel of Fig \ref{fig:AwschalomMCD}). Ando \textit{et al.} also demonstrated that the broad positive "background" in the MCD signal extends well below the fundamental band gap of GaAs ($\omega<0.6$ eV), from which they concluded that this positive signal must originate from the Mn induced impurity band.\cite{ando-2007,ando:067204} This assertion has recently been confirmed by theoritical calculations of the MCD signal resulting the Mn induced impurity band.\cite{tang-2008} 
	 
	The MCD results therefore appear to be in concert with the other techniques that suggest the ferromagnetism in Ga$_{1-x}$Mn$_{x}$As is mediated by holes in a Mn induced impurity band, containing a significant portion of the Mn d-states. Specifically, if the impurity band states originated from the Mn d-shell, then one would expect these states to have a ferromagnetic exchange with the local moment that is much stronger than that of the valence band, as indicated by the large and positive MCD spectra they induce. Furthermore, the lack of a Moss-Burstein shift of the fundamental band gap of Ga$_{1-x}$Mn$_{x}$As supports the notion of holes in the impurity band and not the valence band. Lastly, since the impurity band results from atoms that are somewhat randomly placed in the lattice, one might expect that any features they produce would be rather broad in energy. In addition, it should be noted that recent ab-initio calculations also suggest the sign and shape of the MCD spectra are sensitive to the defects in Ga$_{1-x}$Mn$_{x}$As (ie: As$_{Ga}$ and Mn$_{i}$)\cite{picozzi:235207}. Interestingly, X-Ray MCD is also effected by defects in Ga$_{1-x}$Mn$_{x}$As.\cite{kronast:235213,wu:153310} These calculations imply a careful study of the magneto-optical spectra of Ga$_{1-x}$Mn$_{x}$As upon annealing would uncover the intrinsic contributions to the MCD spectra. 

	Before closing this section we would also like to discuss two studies of the magneto-optical properties of Ga$_{1-x}$Mn$_{x}$As/Al$_{y}$Ga$_{1-y}$As quantum wells. In both experiments the barrier height, determined by the Al concentration, was fixed while the width of the Ga$_{1-x}$Mn$_{x}$As layer was varied\cite{OiwaA:Forqsa,ShimizuH:Blumsa}. Typically a clear blue-shifting of the spectra was seen with decreasing well width. This blue-shifting was attributed to the formation of sub-bands due to the confinement of the holes in the Ga$_{1-x}$Mn$_{x}$As layers. Specifically as the well width is reduced the holes are more tightly bound and the sub-bands move lower in energy. This assertion was confirmed by calculations based on the Kronig-Penney model. We also note that the overall strength of the MCD spectra grew with decreased well width, which may result from an enhanced p-d exchange due to the greater confinement of the holes. 

\subsection{Photoluminescence in Ga$_{1-x}$Mn$_{x}$As }
\label{sec:luminescenceGaMnAs}
	In a typical photoluminescence study, electrons are excited via a laser with an energy that is at or above the band gap. These excited electrons then relax towards the zone center of the conduction band via emission of optical and/or acoustic phonons. The excited electrons eventually recombine with holes either in the valence band or defect states, leading to the emission of light at energies below the excitation energy. In general defects in semiconductors tend to have rather large cross sections for recombination, therefore luminescence spectra allow for accurate determination of defect levels. However the defects are also rather good traps for the photo-excited electrons, therefore they can also lead to a quenching of the signal. Nonetheless  photoluminescence has proven to be a powerful technique for determining the electronic structure of doped semiconductors\cite{CardonaBook}. 

\begin{figure}
\center
\includegraphics[width=18pc]{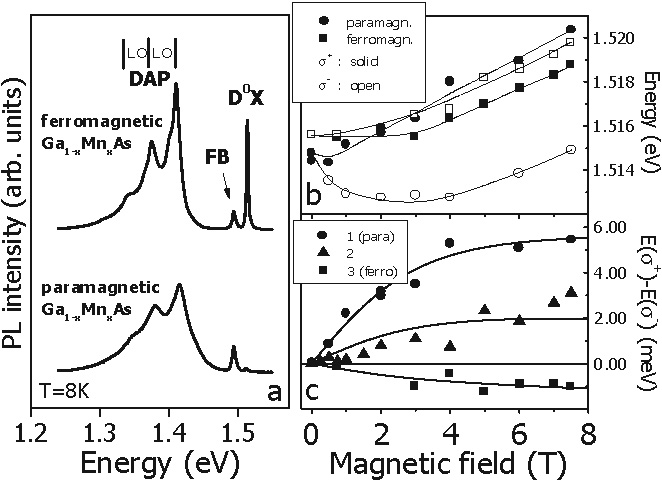}
\caption{\label{fig:GaMnAsPL}(a) Photoluminescence spectra taken from reference \cite{HeimbrodtW:Monsrt}, for a ferromagnetic and paramagnetic Ga$_{1-x}$Mn$_{x}$As layer grown by MOVPE. (b) Energy position of the Pl for left and right circularly polarized light as a function of magnetic field B. (c) Zeeman splitting versus field for the different samples. All spectra are taken at T =2 K.}
\end{figure}
	
	The first studies of the photoluminescence (PL) spectra of Ga$_{1-x}$Mn$_{x}$As focused on samples prepared by metal-organic vapor phase epitaxy (MOVPE). The measured PL spectra shown on the left side of Fig \ref{fig:GaMnAsPL} display two sharp peaks that result from defects in the buffer layer used in the growth, namely a donor to bound exciton transition at 1.514 eV ($D^{0}X$) and a free to bound transition (FB) due to C$_{As}$ at 1.495 eV\cite{HeimbrodtW:Monsrt}. The broad emission lines around 1.41 eV result from inter-band transitions in the Ga$_{1-x}$Mn$_{x}$As layer and display phonon replicas (marked as LO in Fig. \ref{fig:GaMnAsPL}). Due to the optical selection rules, transitions between the conduction and valence bands emit circularly polarized light, whose sense (ie: left vs. right) is related to the spin of the conduction electron and valence band hole involved in the transition. Therefore by monitoring the splitting of emission lines of left and right circularly polarized light, one can determine the Zeeman splitting of the conduction and valence bands. In a non-magnetic semiconductor this splitting ($\Delta E_{PL}$) will be linear in field ($\Delta E_{PL} = g_{exc} \mu_{B}B$), where $g_{exc}$ is the excitonic g-factor. However in a DMS there will be a nonlinear response due to the exchange splitting. Specifically, $\Delta E_{PL} = g_{exc} \mu_{B} B + x_{mn} \cdot N_{0}(\alpha-\beta)<S_{z}>$, where $x_{mn}$ is the percentage of Mn local moments and $<S_{z}>$ is the average value of the magnetization of the Mn moments. By fitting the field dependence of the splitting of $\sigma^{+}/\sigma^{-}$ of Ga$_{1-x}$Mn$_{x}$As, the authors determined $N_{0}\beta=2.5~eV$ in their paramagnetic sample whereas they found $N_{0}\beta=-0.28~eV$ in their ferromagnetic sample. This change in sign with doping appears to be consistent with the MCD results outlined in sub-section \ref{sec:magOptGaMnAs}, and likely results from a change in the origin of the states involved in the emission.

	This work was followed up by measurements of the photoluminescence in Ga$_{1-x}$Mn$_{x}$As/Al$_{0.4}$Ga$_{0.6}$As quantum wells\cite{poggio:235313,PhysRevLett.95.017204}. These samples were carefully prepared such that their defect content was minimized. In fact, Poggio \textit{et al.}\cite{poggio:235313} and Myers \textit{et al.}\cite{PhysRevLett.95.017204} specifically identified a peak in their PL spectra due to Mn$_{i}$, providing a unique opportunity to monitor the electronic contribution and concentration of this important defect. Interestingly the Zeeman splitting of a narrower peak in the PL spectra was found to track the magnetic field and well width (ie: thickness of the Ga$_{1-x}$Mn$_{x}$As layer) expected for heavy holes. Therefore the authors attribute this peak to heavy hole - exciton recombination and attempted to determine $N_{0}\beta$ from its splitting in magnetic field. However they found that the splitting was highly nonlinear with applied magnetic field and samples with larger Mn concentrations could not be fit well. 

\begin{figure}
\center
\includegraphics[width=18pc]{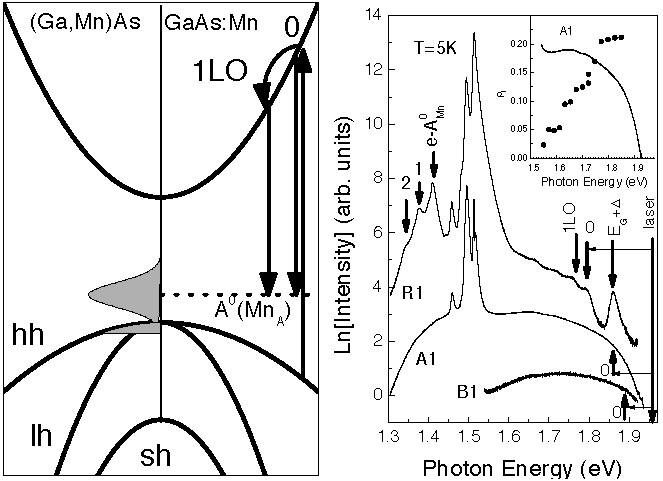}
\caption{\label{fig:HPL_explanation}Left Panel: Schematic of HPL spectroscopy. The vertical arrow on the right shows excitation of the electrons from the valence band to the conduction band. The luminescence transitions from the point of creation and after emission of an LO phonon are indicated by vertical down  arrows. The horizontal dashed line on the right presents mono-energetic distribution of the acceptor states in doping regime. Right Panel: HPL spectra of samples in order of increasing manganese content: R1, A1, and B1, that were excited with He-Ne laser (1.96 eV) at T=5 K. The arrows labeled as Ò0Ó and Ò1LOÓ indicate the energy for recombination of electrons from the point of generation and after emission of one LO phonon, respectively. The recombination of equilibrium electrons with holes bound to Mn acceptor is marked by the arrow $e-A^{0}_{Mn}$. Peaks labeled as Ò1Ó and Ò2Ó are the first and second LO replica of the $e-A^{0}_{Mn}$. The PL
band related to the recombination of holes in the spin-orbit split-off band with the Mn double donor is labeled as $E_{G}+\Delta$ \cite{Ploog_HPL1,Ploog_HPL2}.}
\end{figure}

	In addition to attempts to extract the exchange constants, luminescence studies have also been used to explore the intrinsic electronic structure of Ga$_{1-x}$Mn$_{x}$As. A particularly powerful tool is hot-electron photoluminescence (HPL) where the excitation is well above the fundamental band gap, such that the excited electrons have a kinetic energy ($\frac{\hbar k^2}{2m^{*}}$) much bigger than their thermal energy ($k_{B}T$). The power of this technique is two fold. First, HPL generally leads to a large polarization of the emission, easing the determination of exchange constants. More importantly it provides a unique ability to measure the separation of the impurity level from the valence band. This assertion is explained in the left panel of Fig \ref{fig:HPL_explanation} where a schematic of the HPL process is shown. Specifically, since the electrons have $\frac{\hbar k^2}{2m^{*}}>k_{B}T$ they have a significant probability to recombine with holes at the Fermi level before emitting phonons to relax towards the $\Gamma$ point\cite{Ploog_HPL1,Ploog_HPL2}. Therefore the onset of HPL signal occurs at an energy: $\hbar\omega_{0}=\hbar\omega_{exc}-E_{A}-E_{hh}$, where $\hbar\omega_{exc}$ is the excitation energy, $E_{A}$ is the acceptor binding energy, and $E_{hh}$ is the heavy hole kinetic energy. This technique was recently exploited to uncover the nature of the holes in Ga$_{1-x}$Mn$_{x}$As\cite{Ploog_HPL1,Ploog_HPL2}. As shown in the right panel of Fig  \ref{fig:HPL_explanation}, using a He-Ne (632.8nm) laser to excite their samples, Sapega \textit{et al.} found a significant separation between the onset of the luminescence and  excitation energy. This indicates that the holes reside in an impurity band for concentrations as high $x=0.04$. Furthermore it confirms the infra-red results that indicate a reduction in the separation between the IB and the VB upon an increase in Mn doping levels\cite{Burchannealled}.  

	This powerful technique also provides other important insights into the physics of Ga$_{1-x}$Mn$_{x}$As. Specifically if some holes reside in the valence band due to  thermal excitation, one expects to see a signal at the excitation energy at high temperatures, as was observed in all samples studied. Nonetheless, with increasing doping levels, emission at the excitation energy appeared at smaller temperatures. Thus the energy required for thermal activation, is reduced with an increase in Mn concentration. This confirms the results of the optical conductivity studies that indicate a reduction in the separation between IB and VB with increasing doping concentration (see sub-section \ref{sec:GaMnAsIR}). Furthermore it was found that the polarization of the emitted light was close to that of the excitation laser, indicating that the optical selection rules involved in the recombination at the band gap were not relaxed significantly by disorder. This is consistent with the observation of an MCD signal at the band gap observed in Ga$_{1-x}$Mn$_{x}$As (see Section \ref{sec:magOptGaMnAs}).  
		
	Sapega \textit{et al.} also studied the polarization of the HPL spectra ($\rho_{c}=\frac{I^{+}-I^{-}}{I^{+}+I^{-}}$) as a function of magnetic field in the Faraday geometry\cite{Ploog_HPL1,Ploog_HPL2}. They found that the data for a paramagnetic sample with a Mn concentration of $5\times10^{17}~cm^{-3}$ were well explained by a model wherein the holes are antiferromagnetically coupled to the Mn spins. This behavior was also exhibited by ferromagnetic samples at temperatures well above their respective $T_{C}$. However at low temperatures the ferromagnetic samples displayed a quick onset of polarization with field, but the polarization did not saturate at the highest fields measured (10 T). The authors were able to successfully explain this behavior in a model where the samples were mostly ferromagnetic with about thirty percent of each film remaining paramagnetic. This conclusion is supported by a recent MCD study that also observed ferromagnetic and paramagnetic behavior\cite{furdynaMCD}. However both studies were performed in as-grown films, not in annealed samples. Nonetheless, the saturation values of the polarization were still much smaller in the ferromagnetic samples than what was observed in the paramagnetic samples. The authors suggest this effect originates from the splitting of the impurity band states of different angular momentum. They also argue that such splitting is a natural result of the strain inherent in Ga$_{1-x}$Mn$_{x}$As thin films. This apparent two-component behavior confirms predictions of models in the impurity band limit, where the disorder results in a distribution of exchange values throughout the system\cite{berciu:045202,PhysRevB.65.115308,PhysRevB.67.155201,PhysRevLett.88.247202}. 
	
\subsection{Time Resolved studies of Ga$_{1-x}$Mn$_{x}$As}
	  	
	Given its potential in magneto-optical devices, it is not surprising that the majority of time-resolved studies of Ga$_{1-x}$Mn$_{x}$As have focused on light induced changes in the magnetization (M). The first such study involved a two-color (probe energy = 1.55 eV, pump energy =3.20 eV  ) time resolved Kerr rotation (TRKR), with a pump fluence corresponding to $6.2\times10^{19} cm^{-3}$ injected carriers\cite{kojima:193203,kojima-2004}. We note that in general TRKR measurements are performed with the magnetization in the sample plane ($\overrightarrow{x}$ direction) whereas the pump induces and the probe measures a magnetic moment normal to the plane ($\overrightarrow{z}$ direction). Following this approach in an annealed Ga$_{0.94}$Mn$_{0.06}$As sample with a $T_{C}=110~K$ Kojima \textit{et al.} determined the rotation of the polarization state of the reflected light ($\Delta\Theta$) as well as the relative change in the intensity of the two linear polarizations, often referred to as the ellipticity ($\Delta\eta$)\cite{kojima:193203,kojima-2004}. 
	
\begin{figure}
\center
\includegraphics[width=18pc]{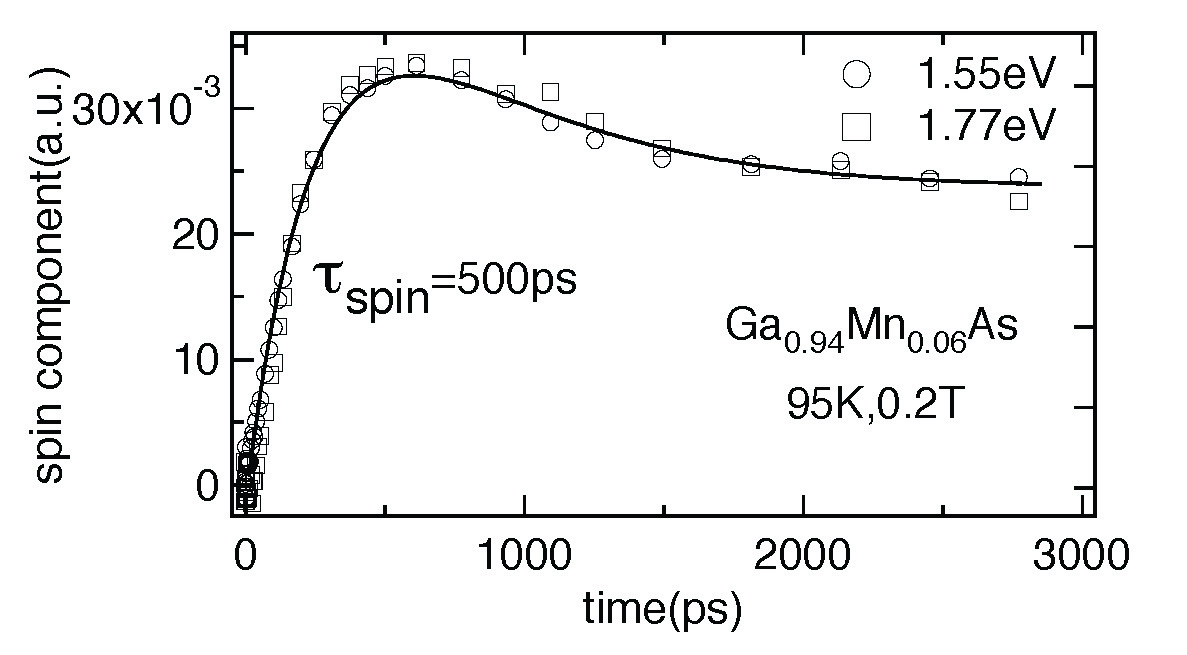}
\caption{\label{fig:Kojima} Time evolution of the extracted magnetization (the open circles were determined from $\Delta\Theta$ and $\Delta\eta$ at 1.55 eV, and the
open squares are from $\Delta\Theta$ and $\Delta\eta$ at 1.77 eV). The data is normalized to their value at the maximum point. A positive signal indicates demagnetization. A solid line shows a result obtained using the three-temperature model considering thermal diffusion\cite{kojima:193203,kojima-2004}. }
\end{figure}

	Since the ellipticity and rotation are both proportional to M, the pump induced changes can be written as:

\begin{equation}
\label{eq:KerrRotation} 
\Delta(\Theta,\eta)\approx f_{(\Theta,\eta)} \Delta M(t) + \Delta f_{\Theta,\eta}M
\end{equation}

where $f_{\Theta ,\eta}$ are functions dependent on the optical constants of the material at the wavelength being probed. Since changes in M are likely to be slow, the initial changes in $\Delta \Theta$ and $\Delta \eta$ are dominated by the second terms in equation \ref{eq:KerrRotation}. In accord with this assertion, Kojima \textit{et al.} found that in the first few picoseconds $C\Delta \Theta=\Delta \eta$, indicating $C \Delta f_{\Theta}=  \Delta f_{\eta}$\cite{kojima:193203,kojima-2004}. Thus one can determine the pump induced change in magnetization via: $(C\Delta f_{\Theta}-  \Delta f_{\eta})\Delta M(t)=C\Delta\theta(t)-\Delta\eta(t)$. The resulting change in magnetization as a function of time is shown in Fig \ref{fig:Kojima}. Similar to In$_{x}$Mn$_{1-x}$As (see sec. \ref{TRInMnAs}), the demagnetization in  Ga$_{x}$Mn$_{1-x}$As occurs over a long time (500 picoseconds). 

While this long time scale could be indicative of a weak coupling between the carriers and the Mn, Kojima \textit{et al.} suggested it results from Ga$_{x}$Mn$_{1-x}$As being half-metallic (ie: only one spin population exists at the Fermi energy) \cite{kojima:193203,kojima-2004}. Specifically if the system is half-metallic, spin flips are difficult and therefore the specific heat of the spin system is rather low. It has recently been shown by Wang et. al., that this long time constant can be reproduced via theoretical calculations if one assumes weak spin-lattice coupling\cite{KonoReview}. In particular, the laser pulse heats the hole and spin baths, which cool by exchanging heat with the lattice. Therefore the phonons must escape from the laser spot before the sample returns to equilibrium. 

	A complimentary analysis of the time resolved magneto-optical spectra of Ga$_{1-x}$Mn$_{x}$As was also employed. Mitsumori \textit{et al.} measured the TRKR spectra for a ``lightly" ($x=0.011$) and ``heavily doped" ($x=0.068$) sample\cite{mitsumori:033203}. In this case the pump and probe were degenerate at 1.579 eV and the change in reflectance ($\frac{\Delta R}{R}$) that resulted from the pump beam was also measured. In both samples, above $T_{C}$ a fast rise in the induced reflectance was observed that exponentially decayed within $\approx25$ picoseconds regardless of the incoming pump polarization. Simultaneously a change in the Kerr signal was seen that also decayed exponentially with a similar time constant. However the sign of the induced change in the Kerr signal depended on the polarization of the incoming light. These results indicate that the change in the Kerr signal is a product of the spin-polarization of the photo-induced carriers. 

\begin{figure}
\center
\includegraphics[width=18pc]{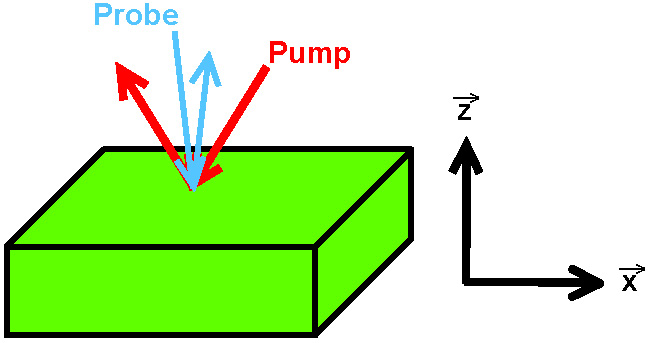}
\caption{\label{fig:ppdiagram} A diagram of a typical pump-probe experiment.}
\end{figure}
	
	Interestingly, below T$_{C}$ a different behavior was detected. In both samples below their respective Curie temperatures, two exponentials were seen in the TRKR signal. The fast exponential in the Kerr signal had a similar time constant as what was observed in $\frac{\Delta R}{R}$, indicating it is due to the thermalization of the excited carriers. Mitsumori \textit{et al.} attribute the long exponential, whose strength grows as the temperature is reduced below $T_{C}$, to the pump induced rotation of Mn spins from $\overrightarrow{x}$  to $\overrightarrow{z}$  followed by their relaxation back to $\overrightarrow{x}$ (see Fig. \ref{fig:ppdiagram})\cite{mitsumori:033203}.  Surprisingly a similar study by Kimel \textit{et al.}, found similar dynamics of the Kerr rotation, however there was no apparent effect of cooling the sample below T$_{C}$\cite{Kimel_PRL04}. These studies suggests that sample preparation (and in particular defects) play an important role in the ultra-fast dynamics of Ga$_{1-x}$Mn$_{x}$As. 
	
	The important role of holes in controlling the ultra-fast response of Ga$_{1-x}$Mn$_{x}$As has recently been conclusively demonstrated. Specifically, Wang \textit{et al.} recently performed a two-color TRKR measurement with the pump well below the gap (2$\mu m \approx 0.62 eV$) while the magnetic state was probed at the fundamental band gap (1.5eV). This approach injects additional holes into the valence band, while the electrons are put into defect states, such that the electrons do not affect the dynamics. Similar to previous results, a fast reduction in the Kerr signal was seen, whose sign changed with applied magnetic field, conclusively demonstrating that demagnetization occurred. Furthermore a 2 nanosecond recovery of the ferromagnetic state was seen. Interestingly, it has also been shown that one can cause a complete reversal of the direction of the in-plane magnetization via a 100 fs optical pulse.\cite{astakhov:152506} Finally we note a recent study has exhibited an optically induced ferromagnetic, and observed effects that cannot be simply attributed to heating.\cite{wang:217401} 
	
	Cywi\'nski and Sham have explained the demagnetization in Ga$_{1-x}$Mn$_{x}$As and In$_{1-x}$Mn$_{x}$As via a transfer of angular momenta from the local spins to the holes\cite{ShamDemag}. They suggest that an increase in the temperature of the carriers at the Fermi surface by the pump laser allows the local moments to transfer angular momenta to the carriers via the p-d exchange. This process continues until the holes cool via emission of phonons. In addition Cywi\'nski and Sham have connected the band structure and the observed demagnetization in III-Mn-V DMS,\cite{ShamDemag} providing an excellent opportunity to use these materials in ultra-fast magneto-optical devices.  

\begin{figure}
\center
\includegraphics[width=18pc]{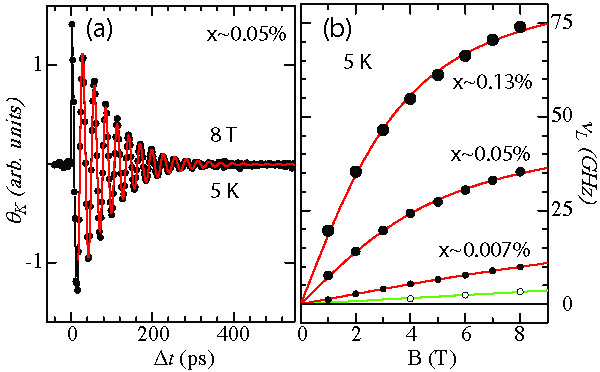}
\caption{\label{fig:GaMnAsTRKR} Time-resolved electron spin dynamics in a Ga$_{1-x}Mn_{x}As/Al_{0.4}Ga_{0.6}As$ quantum well $7.5~nm$ thick. (a) TRKR data (points) together with a fit (red line). (b)  Larmor precession ($\nu_{L}$) as a function of applied magnetic field (B) for different x values (solid points). Open points are the x = 0 sample. Red lines in (b) are fits to eq. \ref{eq:larmor}\cite{poggio:235313}.}
\end{figure}
			
	Time resolved Kerr rotation measurements have also been exploited to determine the strength of the s-d exchange in Ga$_{1-x}$Mn$_{x}$As. Specifically Poggio \textit{et al.}\cite{poggio:235313} and Myers  \textit{et al.}\cite{PhysRevLett.95.017204} focused on Ga$_{1-x}Mn_{x}As/Al_{0.4}Ga_{0.6}As$ quantum wells grown at higher temperatures (ie: $\approx400$ K) with dilute Mn concentrations ($x\leq0.0013$). This resulted in samples with low defect concentrations, enabling careful study of the precession of the photo-excited spins. In these experiments the electrons are injected with their spins in the $\overrightarrow{x}$ direction, while the Mn spins lie along the $\overrightarrow{z}$ axis (due to a small external field B, see Fig. \ref{fig:ppdiagram}). Thus the photo-excited spins rotate around the total magnetic field, such that:
	\begin{equation}
	\label{eq:precess}
	\Theta(t)=Ae^{-t/T_{2}^{*}}cos(2\pi \nu_{L}t+\phi),
	\end{equation}
	 where A is proportional to the total spin injected, $T_{2}^{*}$ is the  in-homogeneous transverse spin lifetime, $\nu_{L}$ electron spin precession (Larmor) frequency, and $\phi$ is a phase factor. Since the Mn is so dilute and the hole spin relaxation time in GaAs is relatively fast, the observed precession can be assigned explicitly to the injected electrons.\cite{poggio:235313,PhysRevLett.95.017204}  
	 
	 The measured Kerr rotation and a fit of the data to eq. \ref{eq:precess} are shown in the left panel of Fig \ref{fig:GaMnAsTRKR}. Since the spins rotate around the combination of the internal (exchange) and external magnetic fields the Larmor precession is: 
	 \begin{equation}
	 \label{eq:larmor}
	 h\nu_{L}=g_{e}\mu_{B}B-xN_{0}\alpha<S_{z}>,
	 \end{equation}
By describing $<S_{z}>$ with a Brillouin function and using the electron g-factor from the $x=0.0$ samples, the value of x$N_{0}\alpha$ was extracted from fits of $h\nu_{L}$ as a function of magnetic field and temperature (see right panel of Fig \ref{fig:GaMnAsTRKR}). The x$N_{0}\alpha$ for samples with various well widths were linear in x, demonstrating the reliability of the technique. Furthermore the value of $N_{0}\alpha$ decreased with increasing well width due to the de-confinement of the electrons. From these data Poggio \textit{et al.}\cite{poggio:235313} and Myers \textit{et al.}\cite{PhysRevLett.95.017204} found $N_{0}\alpha=-0.09~eV$ in bulk  Ga$_{1-x}$Mn$_{x}$As. This suggests the exchange between electrons and Mn in Ga$_{1-x}$Mn$_{x}$As has the opposite sign as what is seen in II-VI DMS\cite{furdyna_review}. Recently, a similar study examined the role of kinetic exchange and confinement in determining $N_{0}\alpha$\cite{stern:045329}.

Time resolved studies of Ga$_{1-x}$Mn$_{x}$As have not been limited to low temperature magneto-optical studies. Yee \textit{et al.} studied samples with $0.008\leq x \leq 0.059$, grown on ZnSe such that the substrate could be removed\cite{yee:113509}. The change in absorption after optical excitation (with approximately 10$^{17}~cm^{-3}$ photo-generated carriers) was then measured as a function of time. In LT-GaAs a reduction in absorption was observed after photo-excitation due to the band filling effect (ie: reduction in possible transitions due to occupation of states from photo generated carriers). Interestingly, in LT-GaAs two relaxation times were observed due to thermalization with the lattice and trapping of carriers by defects. In Ga$_{1-x}$Mn$_{x}$As an increase in absorption occurred over a few hundred fs, that then relaxed exponentially. This was explained via the trapping of photo-excited carriers via defects close to the band edges, such that these additional carriers could then contribute to the absorption near the band edge. The effect of band gap re-normalization was also excluded since the increase in absorption did not depend on the probe wavelength. 

\subsection{Origin of the states near the Fermi level in Ga$_{1-x}$Mn$_{x}$As}
\label{sec:controversy}

An essential issue for descriptions of the properties of Ga$_{1-x}$Mn$_{x}$As is whether the states at the Fermi energy in metallic samples are best described in terms of valence band holes of the GaAs host or if these states preserve impurity band character. A resolution of the issue is needed for the in-depth understanding of carrier mediated ferromagnetism in this prototypical III-V ferromagnet and also for assessment of its potential for spintronics device concepts. In order to resolve this issue, systematic transport and spectroscopic experiments have been carried out by many research teams worldwide and we believe emerging trends overviewed in sub-sections \ref{sec:GaMnAsIR}-\ref{sec:luminescenceGaMnAs} allow one to draw a number of firm conclusions. It should be noted that the nature of electronic transport in heavily disordered and compensated semiconductors is an extremely complicated problem even when all relevant impurities are non-magnetic\cite{MottBook,efrosBook,ThomasReviewMIT,Paalanen:1991ef}. While a cursory survey of the literature on Ga$_{1-x}$Mn$_{x}$As may suggest strong disagreements in the community on Òthe impurity band versus valence band controversyÓ, a deeper inspection of published works uncovers emerging consensus on many substantive matters. 

First, it is uniformly agreed that the properties of Ga$_{1-x}$Mn$_{x}$As in the immediate vicinity of the metal-to-insulator transition (ie: $x\simeq0.02$) are best understood within the impurity band scenario. Importantly, ferromagnetism is registered on {\it both} the insulating and the metallic side of the transition.\cite{VanEsch,SatohY:Carcde} The survival of ferromagnetism in the insulating samples clearly shows that band like transport of holes in valence band is not necessary to produce long range spin order in the system. Second, proponents of either of the two points of view appear to agree on the fact that in metallic samples the valence band states overlap with the impurity band states in the low doping regime.\cite{mahadevan:115211,alvarez:045202,SanvitoGaMnAsPrb,majidi:115205,fiete:045212,MillisPRLGaMnAs,rader:075202,burch:125340,Fujimori_PRB_GaMnAs,Okabayashi:2001lr,jungwirth:PRB,PhysRevB.65.115308,VanEsch,mahieu:712,thomas:082106,russo:033308,PhysRevB.67.205201,PhysRevLett.91.057202,popescu:075206,moca-2007} In this regime, a canonical form of activated electronic transport is neither expected theoretically nor is it experimentally observed.\cite{efrosBook,ThomasReviewMIT} At the same time many features of the dc transport and magneto-transport data in Ga$_{1-x}$Mn$_{x}$As point to deviations from the Fermi liquid picture of mobile holes in the valence band. For example, the latter picture is not consistent with a non-monotonic behavior of the resistivity both in the limit of the lowest T and in the vicinity of the Curie temperature.\cite{hwang:035210,russo:033308,moca-2007,he:162506,rokhinson:161201}

The totality of spectroscopic data discussed in subsections \ref{sec:GaMnAsIR}-\ref{sec:luminescenceGaMnAs} support the view that the IB states preserve their identity derived from the d-character of Mn dopants even in the regime in which these states overlap with the VB on the metallic side of the metal-insulator transition. Conventional wisdom of heavily doped semiconductors teaches us that on the metallic side of the transition the impurity band and the host band merge into one inseparable band (see section \ref{sec:magMIT}). It is worth pointing out that this picture has been developed primarily for non-magnetic dopants and is at variance with the rich physics of resonant states formed by magnetic impurities in metals. Remarkable manifestations of these resonant impurity states include various anomalies of the resistivity as well as a dramatic enhancement of the effective mass of mobile charges compared to the band structure values.\cite{Hewson:1997xz} Furthermore, the optical effective masses in Ga$_{1-x}$Mn$_{x}$As, inferred from the oscillator strength analysis of infrared data (see section \ref{sec:GaMnAsIR}), yield m$^{*}$ on the order of 10 times $m_{e}$\cite{singley:165204,Burchannealled,singley:097203}. These high masses should be contrasted with the value $\sim0.38~m_{e}$ determined from transport of holes in the valence band\cite{songprakob:171}. The above experiments and many additional findings reviewed in this paper are therefore best described within the framework of the Mn induced impurity band formed within the band gap of the GaAs host. 

Holes in Ga$_{1-x}$Mn$_{x}$As uncover an intriguing dichotomy: finite values of the dc conductivity in the limit of T$\rightarrow 0$ unequivocally establish metallic transport due to extended states and yet other properties including a mid-IR resonance in the optical conductivity measurements are best understood in terms of transitions involving the bound impurity states. In addition, the dc transport is quite exotic and is characterized by much lower mobility then in In$_{1-x}$Mn$_{x}$As (see references \cite{IyeY:Mettam,Burchannealled,moca-2007}) along with weak localization effects\cite{rokhinson:161201}. Ohno and Dietl\cite{ohno-2007} have discussed this rather exotic character of III-Mn-V systems in the context of the so-called "two-fluid model" of transport in conventional disordered semiconductors intended to reconcile the coexistence of band-like and impurity-like properties.\cite{Paalanen:1991ef} While such a model is an intriguing possibility, its realization requires nearly phase separated character of the studied materials near the MIT boundary. Furthermore, the materials should become more homogeneous as the carrier density is increased (as the samples become more metallic).  Nonetheless the experimental findings reported here suggest the electronic behavior of samples with the highest T$_C$ values reported to date are best understood in the impurity band scenario. In addition, no local probes have uncovered such effects in Ga$_{1-x}$Mn$_{x}$As.\cite{yakunin:216806,mahieu:712,gleason:011911,PhysRevB.68.235324,grandidier:4001,tsuruoka:2800,mikkelsen:4660} We therefore believe that holes at the Fermi energy in ferromagnetic and metallic  Ga$_{1-x}$Mn$_{x}$As are likely to be described by the impurity band scenario. Nonetheless, further optical experiments in samples where the disorder, Mn and carrier densities are independently tuned would help to solve this controversy. 

\section{Ga$_{1-x}$Mn$_{x}$P}
	Transition metals in GaP have been the subject of extensive studies that examined the evolution of the electronic structure with different TM (see \cite{ZungerGaP} and references therein). Interestingly in dilute Ga$_{1-x}$Mn$_{x}$P the Lyman series corresponding to acceptors bound to the Mn have a large binding energy (400 meV)\cite{ZungerGaP,tarhan:195202}. This suggests the Mn acceptor state is both unconventional and that a metallic state will be exceedingly difficult to produce. In fact, there have been no reports of a metallic state in Ga$_{1-x}$Mn$_{x}$P. Nonetheless, due to its close lattice matching with Si, Ga$_{1-x}$Mn$_{x}$P may turn out to be a particularly fruitful avenue for incorporating DMS in well established semiconductor technologies\cite{pearton:1}. Early reports of growth by molecular beam epitaxy and ion-implantation \cite{overberg:969} suggested a magnetic state in Ga$_{1-x}$Mn$_{x}$P with a T$_{C}$ exceeding room temperature, however these reports remain unconfirmed and did not establish the nature of the carriers mediating the ferromagnetic state. Recently Ga$_{1-x}$Mn$_{x}$P has been prepared by ion-implantation with a maximum $x\approx0.06$ and T$_{C}\approx 60~ K$. This study found that the samples were always insulating and that the Curie temperature was reduced by co-doping with the donor Te\cite{scarpulla:207204}. 

\begin{figure}
\center
\includegraphics[width=18pc]{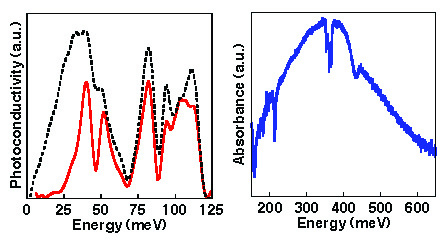}
\caption{\label{fig:GaMnPhotoConductivity} Left Panel: (dashed line) incident intensity, (red lines) photoconductivity of a Ga$_{1-x}$Mn$_{x}$P sample. Right Panel: 10 K absorption spectra of Ga$_{1-x}$Mn$_{x}$P sample compensated with Te . Sharp features in the absorption are due to GaP phonon features\cite{scarpulla:207204}. }
\end{figure}
	
	In order to establish the nature of the bound holes in Ga$_{1-x}$Mn$_{x}$P, Scarpulla \textit{et al.} performed optical absorption and photoconductivity measurements\cite{scarpulla:207204}. The absorption spectra of a Ga$_{0.94}$Mn$_{0.06}$P:Te sample is shown in Fig \ref{fig:GaMnPhotoConductivity}, where a peak is observed centered near 400 meV, the location of the Mn impurity level. The absence of other spectral features suggests the Fermi level is in the Mn impurity band and has not crossed into the GaP valence band. To further confirm that the holes reside in the Mn induced impurity band, the authors measured the d.c. conductivity induced by optical excitation at low temperatures. In Fig \ref{fig:GaMnPhotoConductivity} this photoconductivity spectrum for the Ga$_{0.94}$Mn$_{0.06}$P sample is displayed along with the spectrum for the incident light. The onset for the photoconductivity data is clearly higher in energy than the incident light, indicating the holes reside in bound states that are separated by roughly $\approx26~meV$ from the conducting states in the valence band. This value corresponds well with the size of the gap determined from the thermal activation energy of the resistivity ($31~meV$). By co-doping with the donor Te, the Fermi energy was moved deeper into the Mn induced impurity band and the gap increased\cite{scarpulla:207204}. Lastly, the size of the gap between the Fermi energy and extended states ($26~meV \approx 300K$) indicates that localized holes mediate the ferromagnetism. 
	
\section{Ga$_{1-x}$Mn$_{x}$N}
	For the last decade there has been growing interest in wide band gap, group III-nitrides, for their potential use in high power/temperature electronics, solar-blind UV detectors and blue/green/UV light emitting diodes\cite{pearton:1}. In addition, Dietl \textit{et al.} suggested Ga$_{1-x}$Mn$_{x}$N should posses a $T_{C}$ well exceeding room temperature based upon a mean field model of ferromagnetism in DMS\cite{DietlT:Zenmdf}. This has led to a flurry of research on Ga$_{1-x}$Mn$_{x}$N resulting in its growth by multiple methods including plasma assisted MBE,\cite{edmonds:152114,FerrandD:Spicei,graf:5159,polyakov:4989} metal organic vapor phase epitaxy (MOVPE),\cite{KorotkovRY:OptsGM,KorotkovRY:MnraaP} nebulized spray pyrolysis\cite{SardarK:Magoat}, ion-implantation,\cite{shon:1845} the ammonothermal method\cite{zajac:1276,zajac:4715}, and bulk crystals grown by the high pressure technique\cite{wolos:115210}. In many of these reports ferromagnetic behavior has been claimed to exist at or above room temperature along with the corresponding signatures of a carrier mediated magnetic state, such as the anomalous Hall effect\cite{pearton:1}. However others have either reported ferromagnetism at only very low temperatures,\cite{edmonds:152114} or have provided evidence that the high $T_{C}$ ferromagnetism originates from clusters\cite{Ando:2003Ir,graf:9697,shon:1845,zajac:4715}. Therefore it is not surprising that to date there is no consensus on whether ferromagnetism in Ga$_{1-x}$Mn$_{x}$N is intrinsic or due to a secondary phase in the GaN host. Nonetheless, a consensus has emerged regarding the effects of the Mn dopant on the GaN band structure. Therefore our discussion will focus on studies elucidating the changes in the GaN band structure due to the Mn impurity.
		
		Before discussing the optical properties, we address a key materials growth issue that has emerged in the course of attempts to produce ferromagnetic Ga$_{1-x}$Mn$_{x}$N. Due to the lack of commercially available GaN substrates, Ga$_{1-x}$Mn$_{x}$N is often grown on lattice mismatched substrates (e.g. SiC, Si, and saphire). While this is advantageous for the incorporation of GaN-based devices into standard electronics, it also results in a large dislocation density. Therefore as-grown GaN generally has a very high number of charged defects resulting in a very high electron density\cite{pearton:1}. As a result, Ga$_{1-x}$Mn$_{x}$N is usually found to be n-type and must be co-doped with Mg to be made p-type. 	
				
\subsection{Absorption spectra of Ga$_{1-x}$Mn$_{x}$N}
\begin{figure}
\center
\includegraphics[width=18pc]{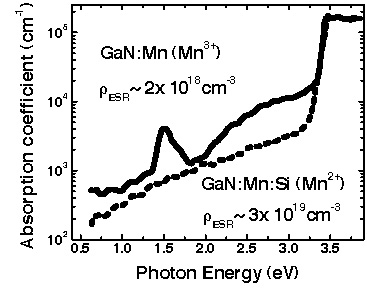}
\caption{\label{fig:GaMnNabsorption} Optical absorption of Ga$_{1-x}$Mn$_{x}$N samples before and after Si co-doping. Only the Ga$_{1-x}$Mn$_{x}$N sample reveals features at 1.5 eV and above 1.8 eV, indicating E$_{F}$ is above the Mn d-levels in Ga$_{1-x}$Mn$_{x}$N:Si\cite{graf:5159}.}
\end{figure}

	Some of the earliest data indicating that Mn formed a deep acceptor level in Ga$_{1-x}$Mn$_{x}$N were optical absorption spectra\cite{KorotkovRY:OptsGM,KorotkovRY:MnraaP,Noh}.  A typical spectrum is shown in Fig \ref{fig:GaMnNabsorption}, where two prominent features are observed, one centered at 1.5 eV and the other with an onset around 2.0 eV. From these spectra it was initially believed that Mn produced an acceptor level approximately 1.4 eV above the valence band of Ga$_{1-x}$Mn$_{x}$N. Specifically, the lower energy peak was assigned to optical transitions of electrons from the valence band to the Mn acceptor level. The higher energy resonance was then assigned to transitions from the acceptor level to the Ga$_{1-x}$Mn$_{x}$N conduction band\cite{KorotkovRY:OptsGM,KorotkovRY:MnraaP,korotkov:1731}. Support for the notion that the holes in Ga$_{1-x}$Mn$_{x}$N do not reside in the valence band was also provided by an extensive study of the band gap. In particular it was shown that the small changes in the band gap could be attributed to structural changes from the Mn and not the Moss-Burstein effect\cite{Zhang_Absorption_GaMnN}.
		
	It was quickly realized that the original assignment of the 1.4 eV and 2.0 eV features did not explain the wealth of data that emerged. In particular, if the assignment was correct, one might expect that adding electrons to the system would fill the Mn impurity band, such that the lower energy transition (from the valence band), would no longer be possible. At the same time one would also expect that adding donors to the system would also result in an increase in the spectral weight of the higher energy transition. However, it was soon shown that co-doping Ga$_{1-x}$Mn$_{x}$N with Si (ie: adding electrons), resulted in a removal of spectral weight from both features (see Fig \ref{fig:GaMnNabsorption})\cite{graf:5159}. Furthermore, spectra taken at low temperatures revealed a significant sharpening of the lower energy feature and not the higher energy resonance. In fact, at low temperatures one can resolve phonon-replicas of the lower energy resonance (ie: resonances that involve the main transition plus the emission of an optical phonon). This indicates that the two features have significantly different origins. 
		
\begin{figure}
\center
\includegraphics[width=18pc]{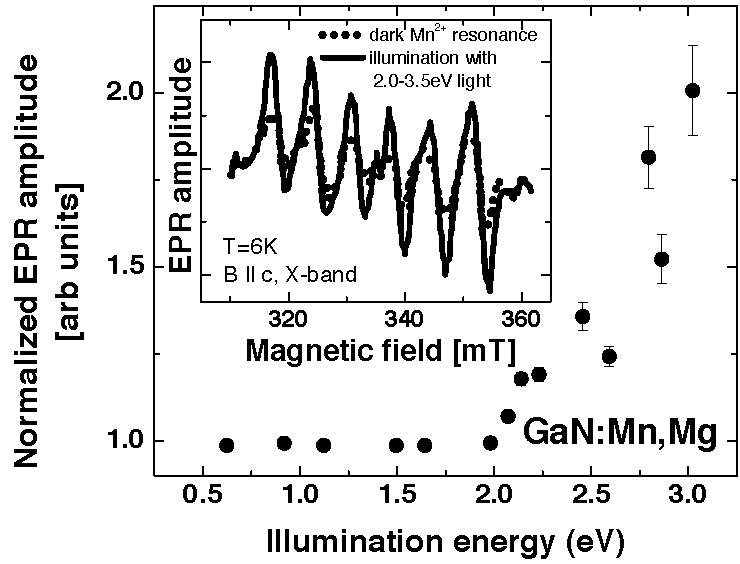}
\caption{\label{fig:GaMnNEPR} Inset: EPR spectra of  Mn$^{2+}$ (d$^{5}$) Ga$_{1-x}$Mn$_{x}$N before (dashed Lines)  and after (solid lines) illumination at 2.0 eV. Main Panel: Strength of the EPR  (d$^{5}$) signal as a function of illumination energy, indicating the charge state of the Mn only changes for light greater than 2.0 eV\cite{wolos:115210}.}
\end{figure}

		An alternate interpretation of the data soon emerged. Namely that the Mn d-levels lie in the gap of GaN with the upper most $E_{G}$ level approximately 2.1 eV above the VB. Therefore the 2.0 eV feature results from transitions between the valence band and the Mn $^5E$ level, while the lower energy absorption was attributed to internal Mn transitions between the $^5T_{2}$ and $^5E$ levels\cite{graf:5159,graf:9697}. Such an assignment helps to explain a number of the results described above. Specifically, the sharpness of the low energy feature results from the fact that it involves transitions between two narrow bands with much smaller bandwidth than the valence or conduction bands. This scenario also explains the evolution of both resonances upon adding electrons to the system.
		Electron paramagnetic resonance experiments (EPR) have also helped to resolve the origin of that these features. Specifically, EPR experiments revealed that the Mn induced absorption features were only observed in samples wherein the Mn was in a 3+ charge state (ie: $d^{4}$) and disappeared in samples where the Mn was in a 2+ (ie: $d^5$) configuration\cite{graf:5159,graf:9697}. Additionally, an increase in the EPR signal of the Mn$^{2+}~(d^{5})$ was observed when the samples were illuminated with light\cite{wolos:115210}. However, as shown in Fig \ref{fig:GaMnNEPR}, this effect only occurred for light with energies above 2.0 eV, indicating that the absorption at lower energies did not involve a change in the charge state of the Mn. Therefore the EPR results confirm the assertion that the optical absorption observed below 2.0 eV only involves transitions on the Mn atom. Whereas for absorption above 2.0 eV, the transitions result in an increase in the number of electrons on the Mn atom, which is consistent with a transition from the N valence band to the Mn d-level.
		
\subsection{Magneto-Optics of Ga$_{1-x}$Mn$_{x}$N}	
\label{sec:GaMnNMagOpt}
	A number of groups have studied the magneto optical response of Ga$_{1-x}$Mn$_{x}$N using a variety of techniques. The earliest studies of the magneto circular dichroism of Ga$_{1-x}$Mn$_{x}$N were done by K. Ando\cite{Ando:2003Ir} wherein no ferromagnetic component was observed. This strongly suggested that the room temperature ferromagnetism is not an intrinsic property of the Ga$_{1-x}$Mn$_{x}$N samples studied. Later, work by D. Ferrand \textit{et al.}, did register an MCD signal from which they extracted a Zeeman splitting that was proportional to the magnetization at low temperatures. They were therefore able to extract a measure of the exchange constants: $N_{0}(\alpha-\beta)=0.2 eV$\cite{FerrandD:Spicei}. While this value is low, one should note that the MCD features were quite broad, and at the band gap two features were overlapped. Nonetheless, the fact that the low temperature MCD signal was observed at the fundamental band gap of GaN indicates that the low temperature ferromagnetism is indeed intrinsic and not due to the formation of second phase.\cite{KojiAndo06302006}

\begin{figure}
\center
\includegraphics[width=18pc]{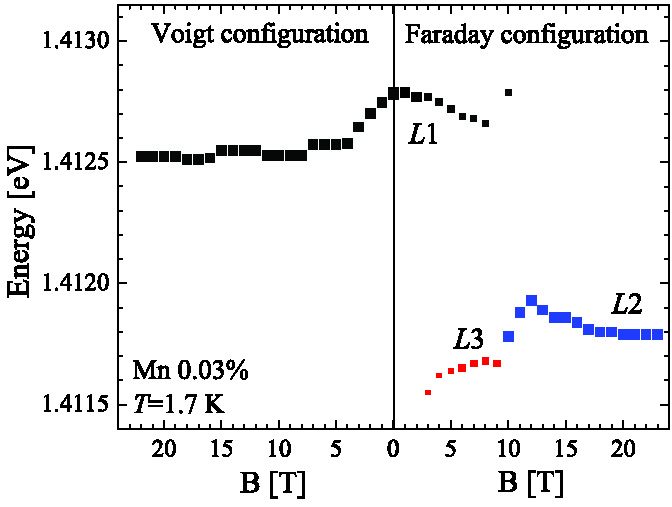}
\caption{\label{fig:Ferrandpositions} Center Energy of absorption features in a Ga$_{0.97}$Mn$_{0.03}$N layer as a function of the magnetic field in Faraday and the Voigt configurations. The symbol size is proportional to the intensity of the feature\cite{marcet-2006-}.}
\end{figure}
	
	In addition to the MCD work mentioned above, two groups have also measured the change in absorption with applied magnetic field. The first such study by A. Wolos \textit{et al.}, focused on the effects of magnetic field on the 1.4 eV feature, where small shifts in the zero-phonon line (ZPL) were seen with applied magnetic field. However at 7 T a sudden jump occurred, after which a gradual shift of the peak continued. Interestingly little effect of magnetic field was observed in the Voight configuration. This behavior is completely at odds with what one would expect if the transition originated in the valence band, providing further evidence that the 1.4 eV feature results from Mn d to d transitions. To confirm this assignment Wolos \textit{et al.} solved a model for the Mn levels that accounted for crystal field splitting, a static Jahn-Teller distortion, the trigonal crystal field, the spin-orbit interaction and the Zeeman splitting of the levels\cite{wolos:245202}. In this model one assumes that the ZPL results from transitions between $^5T_{2}$ and $^5E$ states. From spin-orbit coupling, the states can then be labeled with magnetic numbers $m_{s}=2,1,0,-1,2$, the energies of which will have different field dependencies. Therefore the calculation suggests that at small fields the shifts seen in the absorption are primarily due to the Zeeman effect. However at larger fields two of the levels cross, leading to a jump in the energy of the resonance\cite{wolos:245202}.

\begin{figure}
\center
\includegraphics[width=18pc]{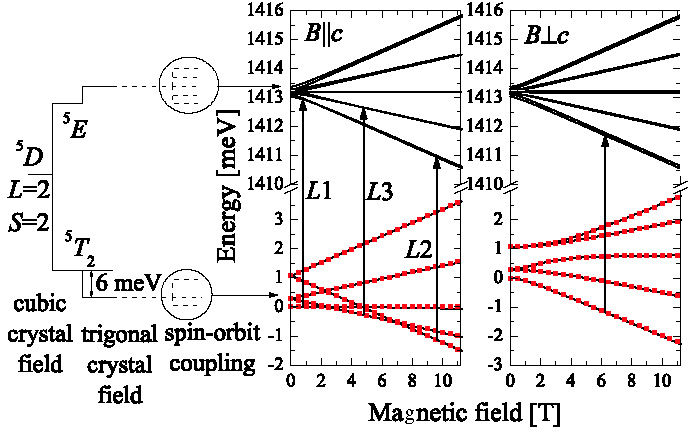}
\caption{\label{fig:marcetlevels} Calculated energy levels including the cubic and trigonal crystal field, spin-orbit coupling, the Zeeman and Jahn-Teller effects. Arrows indicate the origin of the transitions whose positions are plotted in Fig \ref{fig:Ferrandpositions}\cite{marcet-2006-}.}
\end{figure}
	
	 The hypothesis of the dominant role of internal Mn transitions was later confirmed by S. Marcet \textit{et al.}, who studied the magneto-optical absorption at much lower dopings, such that the disorder was minimized\cite{marcet-2006-}. From their spectra, one clearly observes three peaks in the absorption that only exist at certain magnetic field values and orientations. The different peak positions are shown in Fig \ref{fig:Ferrandpositions}, indicating that the apparent jump in the spectra observed in the earlier study\cite{wolos:245202} is due to a shift in spectral weight from the L1 to L2 transitions. In Fig \ref{fig:marcetlevels} we show the relevant energy level scheme as calculated  by Marcet \textit{et al.}, that explains the origin of the three transitions and their field dependence\cite{marcet-2006-}.
		
\subsection{Luminescence of Ga$_{1-x}$Mn$_{x}$N}	
\label{sec:GaMnNLum}

\begin{figure}
\center
\includegraphics[width=18pc]{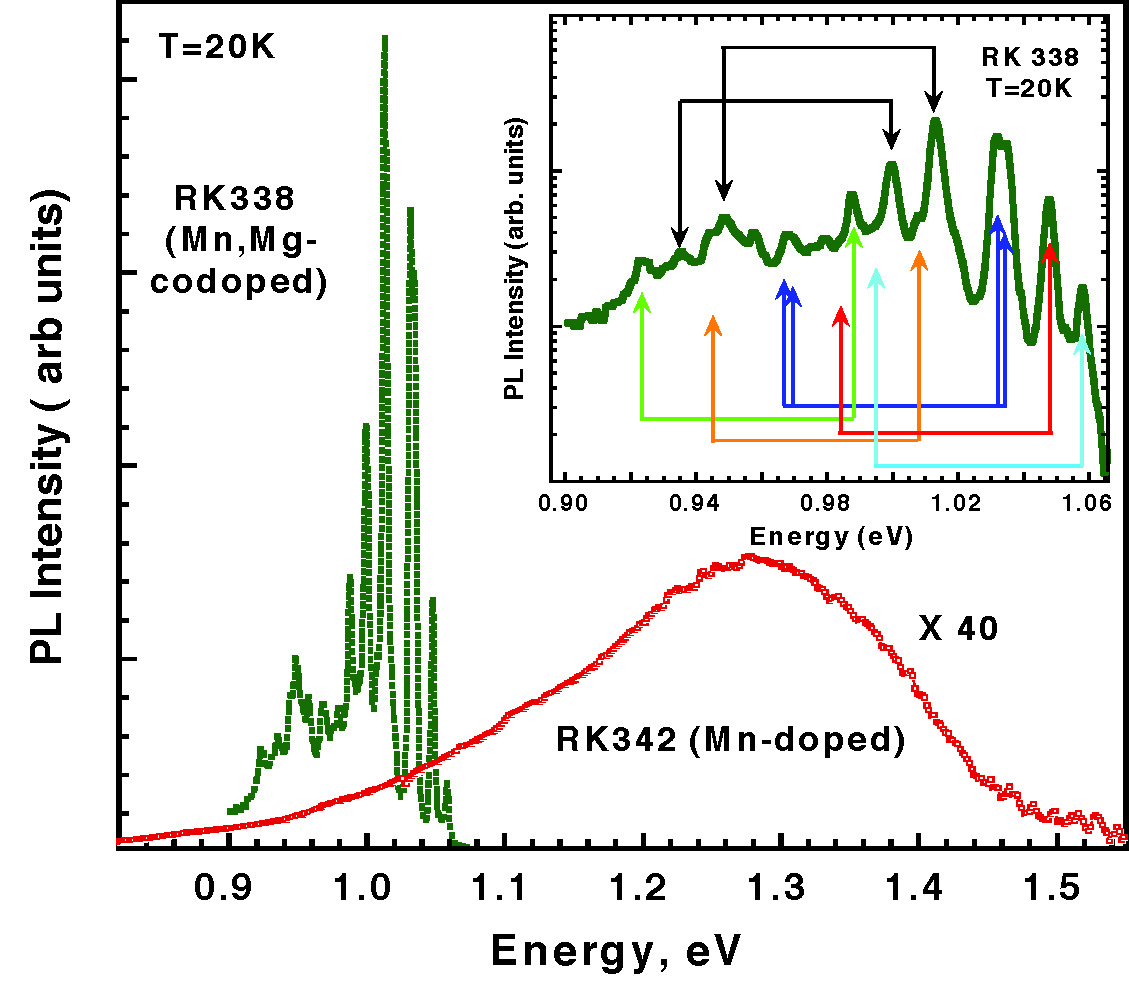}
\caption{\label{fig:wesselsPL} Photoluminescence spectra for Ga$_{1-x}$Mn$_{x}$N before and after doping with Mg. Inset: the sharp peaks in the Ga$_{1-x}$Mn$_{x}$N:Mg with arrows indicating phonon replicas. After doping with Mg, the spectra move to lower energies and sharpen substantially. This indicates the Fermi energy is above the Mn d-level in as-grown Ga$_{1-x}$Mn$_{x}$N, with excitations originating between the d-level and the valence band. Upon doping with Mg, the Fermi energy moves into the Mn levels resulting in an intra-ion transition\cite{han:5320}.}
\end{figure}

	As described above luminescence studies can be quite powerful due to their ability to investigate defect levels and the well-defined selection rules involved in the emission of light. We first focus on the mid-gap photoluminescence, performed by Korotkov \textit{et al.}, wherein regardless of the excitation wavelength a feature with a number of replicas was seen around 1.0 eV\cite{KorotkovRY:MnraaP,KorotkovRY:OptsGM}. Interestingly this feature also appeared to be rather long-lived (ie: exponential decay with characteristic time: $\tau\approx75~\mu s$). It has been shown that in as-grown Ga$_{1-x}$Mn$_{x}$N a strong, albeit broad emission was seen at 1.25 eV. \cite{han:5320} Upon co-doping with Mg this feature moved to 1.0 eV and sharpened significantly to reveal a number of phonon-replicas (see Fig \ref{fig:wesselsPL}). It was also found that the 1.0 eV feature and its replicas were temperature independent\cite{han:5320}. Luminescence spectra under excitation via electron bombardment have also been measured\cite{polyakov:4989}. These studies revealed a feature around 1.9 eV whose intensity grew with Mn concentration, however no resonances near 1.3 eV were observed. In these later studies, the defect levels and in turn Fermi energies are likely to be different since the samples were grown by MBE and not MOCVD.

	Taken together these results provide additional evidence that Mn levels in Ga$_{1-x}$Mn$_{x}$N lie in the fundamental band gap of the GaN host. Specifically, if one places $E_{F}$ at or just above the Mn $^5E$ level, then one would expect a broad emission spectra due the recombination of electron-hole pairs in the Mn level and the valence band. However upon p-doping deep into the Mn level (with Mg for instance), this recombination pathway is removed. Therefore in Ga$_{1-x}$Mn$_{x}$N one would expect to observe sharp emission features from the recombination of electron-hole pairs in the Mn $^5T_{2}$ and $^5E$ levels. Furthermore these levels should be temperature independent and exhibit rather long lifetimes. 

\section{Summary and Outlook}
	In this review we have detailed the extensive optical studies of III-Mn-V ferromagnetic semiconductors, establishing the electronic structure of these compounds. In table \ref{TBL} we summarize a number of the properties of these materials as well as the results from the experiments outlined in this review. In III-Mn-V DMS the degree of itineracy and the origin of the states at Fermi energy are sensitive to the III-V host through the strength of the exchange that it selects. For instance, In$_{1-x}$Mn$_{x}$As appears to behave as a conventional doped semiconductor in that the MIT occurs within the InAs valence band at doping levels consistent with a non-magnetic impurity. Furthermore, the carriers in In$_{1-x}$Mn$_{x}$As, are reasonably de-localized exhibiting high mobilities and standard valence band masses (see section \ref{sec:CR}). The evolution of the electronic structure in Ga$_{1-x}$Mn$_{x}$As is somewhat exotic. Specifically, the Mn induced impurity band appears to overlap with states due to the valence band and yet mobile charges retain the impurity band character deep into the metallic state (see sub-section \ref{sec:GaMnAsIR}). One outstanding manifestation of the persistence of the Mn impurity sates in metallic Ga$_{1-x}$Mn$_{x}$As is that this compound reveals holes with effective masses two orders of magnitude larger than in In$_{1-x}$Mn$_{x}$As. Not surprisingly, the dc transport is also somewhat unconventional and is characterized by much lower mobility then in In$_{1-x}$Mn$_{x}$As (see references \cite{IyeY:Mettam,Burchannealled,moca-2007}).  In both the nitride and phosphide compounds the Mn acceptor level is very deep, and for Ga$_{1-x}$Mn$_{x}$N is the Mn E$_{2}$ level. This appears to have prevented the emergence of a metallic state or intrinsic ferromagnetism at temperatures exceeding 60 K in Ga$_{1-x}$Mn$_{x}$P and Ga$_{1-x}$Mn$_{x}$N, despite comparable doping levels as those found in Ga$_{1-x}$Mn$_{x}$As. 

\begin{table*}
\caption{\label{TBL} Results for III-Mn-V DMS: effective mass of conducting holes ($m^{*}$), average effective mass for valence band holes with non-magnetic acceptors ($m_{VB}$), and the acceptor level for the Mn dopant (E$_{A}$). $\dagger$For GaN, the Zinc-Blende structure is used. $\star$The ferromagnetism and intrinsic $T_{C}$ remain controversial, for a complete discussion see \cite{KojiAndo06302006,zajac:4715,pearton:1}.}
\begin{tabular*}{1\textwidth}{@{\extracolsep{\fill}}r|c|c|c|c}
\hline
\hline
&
In$_{1-x}$Mn$_{x}$As&
Ga$_{1-x}$Mn$_{x}$As&
Ga$_{1-x}$Mn$_{x}$P&
Ga$_{1-x}$Mn$_{x}$N$^{\dagger}$\tabularnewline
\hline

a ($\AA$) &
6.06&
5.65&
5.45&
4.52\tabularnewline

E$_{G}$ (eV) &
0.415&
1.519&
2.34&
3.28\tabularnewline

m$_{VB}$ (m$_{e}$) &
0.41&
0.53&
0.83&
1.4\tabularnewline

m$^{*}$ (m$_{e}$) &
0.35$^{a}$&
$\geq10^{b}$&
-&
-\tabularnewline

E$_{A}$ (eV) &
- &
0.112$^{c,d}$&
0.388$^{d}$&
1.8$^{e}$
\tabularnewline

Max. T$_{C}~(K)$&
60$^{f}$&
172$^{g}$, 250$^{h}$&
60$^{i}$&
$\star$\tabularnewline
\hline
\hline
\end{tabular*} 
\begin{tabular*}{1\textwidth}{@{\extracolsep{\fill}}lll}
$^{a}$\cite{InMnAs_ptype_cr}&$^{b}$\cite{Burchannealled,singley:097203,singley:165204}&$^{c}$\cite{PhysRevLett.18.443}\tabularnewline
$^{d}$\cite{tarhan:195202}&$^{e}$\cite{graf:5159,graf:9697}&$^{f}$\cite{OhnoH:Molbea,KonoReview}\tabularnewline
$^{g}$\cite{JungwirthMaxTcGaMnAs}&$^{h}$\cite{nazmul:017201}&$^{i}\cite{scarpulla:207204}$\tabularnewline
\end{tabular*} 
\end{table*}
	
\begin{figure}
\center
\includegraphics[width=18pc]{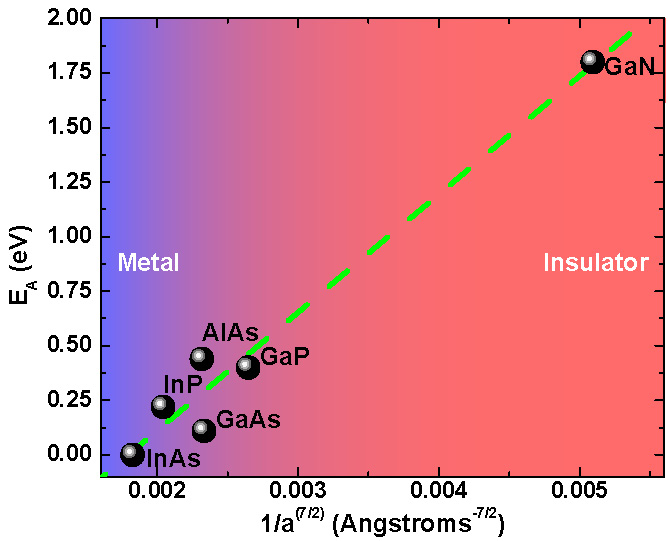}
\caption{\label{fig:Ea} Acceptor level for dilute Mn in III-V compounds ($E_{A}$) versus the inverse lattice spacing ($\frac{1}{a^{7/2}}$). As the hybridization between d and p-orbitals is increased, approximated via $\frac{1}{a^{7/2}}$, the impurity level moves deeper into the band gap, reducing the likelihood of a metallic state. Data taken from \cite{TMLevels} and those indicated in Table \ref{TBL}.}
\end{figure}

	Optical properties of the III-Mn-V series discussed above allow us to draw a number of conclusions. First, producing a higher T$_{C}$ in the III-Mn-V series appears to require a metallic state. Second, generating such a state requires the impurity band to be in proximity to the valence band, otherwise the holes are too strongly localized around the Mn. As suggested theoretically,\cite{majidi:115205,Bhattacharjee,MillisPRLGaMnAs,popescu:075206} this separation is effected by the strength of the hybridization between Mn d electrons and the p-orbitals of its neighbors,  an assertion recently confirmed by a comprehensive study of Ga$_{1-x-y}$Mn$_{x}$Be$_{y}$As and Ga$_{1-x}$Mn$_{x}$As$_{1-y}$P$_{y}$.\cite{alberi:075201} This indicates that this key parameter can be tuned by choosing the appropriate III-V host, a quite novel aspect of these compounds. Indeed, from a simple tight-binding picture, one would expect the hybridization to be inversely proportional to the lattice constant (ie: $V_{pd}\propto\frac{1}{a^{7/2}}$).\cite{Harrison} To explore this possibility in Fig \ref{fig:Ea} we plot the acceptor level in the dilute limit versus $\frac{1}{a^{7/2}}$ for various III-V hosts. As shown in the figure, not only is the electronic structure effected by the strength of $V_{pd}$, as measured by the lattice constant, but this important parameter also controls the degree of itineracy of the holes. 

	Before concluding we would like to make some remarks on the future of this exciting field. The studies outlined here clearly demonstrate a strong connection between the magnetism and optical properties of III-Mn-V semiconductors. They also establish a unique role for the III-V host in combination with a transition metal dopant. Specifically, the trends established here for III-Mn-V DMS could be employed to engineer a desired combination of magnetic, transport and optical properties. Furthermore, a number of available controls including light illumination, electric and/magnetic fields enable manipulation of the interplay between the properties of III-Mn-V semiconductors. Nonetheless the wide array of potential magneto-optical devices these compounds therefore offer, still remains to be fully exploited. 
	
	In addition to the applications of III-Mn-V yet to be implemented, there are also a number of open scientific questions that need to be studied. First and foremost, the nature and strength of the exchange constants in the phosphorus and nitrogen based III-Mn-V semiconductors should be established. Second, a detailed mapping of the broad band (ie: Far and Mid Infrared) magneto-optical response of these compounds may produce useful insights into the nature of the states in the As and P based III-Mn-V DMS, as well as produce magneto-optical devices in this range. Lastly, a realistic picture of the metal to insulator transition in III-Mn-V DMS would help reveal the connections between their magnetic and electronic properties.
	
\section{Acknowledgements}
We are grateful for numerous discussions with O.D. Dubon, M. E. Flatte, J.K. Furdyna, B. Lee, X. Liu, A. MacDonald, A.J. Millis, S. Das Sarma, M.A. Scarpulla, J. Sinova, J.-M. Tang, and C. Timm. We are especially en-debited to S. Crooker, L. Cywinski, M.P. Kennett, and E.J. Singley for careful reading of this manuscript. K.S.B. acknowledges support from the Los Alamos National Laboratory LDRD and the Center for Integrated Nanotechnologies. D.N.B. appreciates support from DOE and ONR award N00014-07-1-347.

\bibliographystyle{elsart-num}
\bibliography{review_bibliography.bib}

\end{document}